\documentclass[12pt]{article}
\usepackage{epsfig}
\usepackage{cite}
\def\GGREPom{{I\!\!P}}
\newcommand{\GGTSindx}[1]{\mbox{\scriptsize #1}}
\newenvironment{GGTSitemize}{\begin{list}{$\bullet$}%
{\setlength{\topsep}{-2.8mm}\setlength{\partopsep}{0.2mm}%
\setlength{\itemsep}{0.2mm}\setlength{\parsep}{0.2mm}}}%
{\end{list}}
\newcounter{GGTSenumct}
\newenvironment{GGTSenumerate}{\begin{list}{\arabic{GGTSenumct}.}%
{\usecounter{GGTSenumct}\setlength{\topsep}{-2.8mm}%
\setlength{\partopsep}{0.2mm}\setlength{\itemsep}{0.2mm}%
\setlength{\parsep}{0.2mm}}}{\end{list}}
\newcommand{\GGTSq}{\mbox{q}}
\newcommand{\GGTSqbar}{\overline{\mbox{q}}}
\newcommand{\GGTSg}{\mbox{g}}
\def\GGpythia{{\sc Pythia}}
\def\GGherwig{{\small HERWIG}}
\def\GGariadne{{\sc Ariadne}}
\def\GGminijet{{\sc Minijet}}
\def\GGphojet{{\sc Phojet}}
\def\GGjimmy{{\sc Jimmy}}
\def\GGgghv01{{\small GGHV}01}
\def\GGggpsone{{\small GGPS}1}
\def\GGggpstwo{{\small GGPS}2}
\def\GGjetset{{\sc Jetset}}
\def\GGtwogam{{\sc Twogam}}
\def\GGtwogen{{\sc Twogen}}
\def\GGdiag36{{\sc Diag36}}
\def\GGpc{{\small PC}}
\def\GGrespro{{\sc Respro}}
\def\GGpdflib{{\small PDFLIB}}
\def\GGj#1{{\it #1}}
\def\GGv#1{{\bf #1}}
\def\GGt#1{``#1''}
\begin{document}
%
\begin{center}
{\LARGE \bf {\boldmath $\gamma\gamma$} Event Generators}
\end{center}
\begin{center}
{\it Conveners}: Leif L\"onnblad, Mike Seymour
\end{center}
\begin{center}
  {\it Working group}: Edouard Boudinov, Jon Butterworth, Ralph
  Engel, Alex Finch, Suen Hou, Maria Kienzle-Focacci, Mark Lehto, Ed
  McKigney, David Miller, Denis Perret-Gallix, Johannes Ranft,
  Gerhard Schuler, Torbj\"orn Sj\"ostrand, Rod Walker, Alison
  Wright.
\end{center}
\vspace*{1.0cm}
\setcounter{tocdepth}{1}
\tableofcontents
\newpage

\section{Introduction}

At LEP~2, two-photon collisions make up by far the largest class of
events.  These are processes in which the incoming electrons each
radiate a photon, which collide to produce a hadronic or leptonic final
state.  A photon can obviously interact electromagnetically with any
charged object.  However, since it has the same quantum numbers as a
vector meson, it can fluctuate into one, and can therefore also be
considered as an incoming hadron, interacting strongly through its
partonic constituents.  The interplay between these two ways of
interacting is unique to the photon and provides much of the interest in
$\gamma\gamma$ physics.

The two-photon invariant mass spectrum is peaked at low mass, so the
bulk of events only produce a few particles, and resonances and
exclusive final states can be studied.  The total cross-section is so
large however, that there are enough events to study deep inelastic
e$\gamma$ scattering and high-$p_\perp$ jets and heavy quark production
in $\gamma\gamma$ collisions.

LEP~2 is the highest energy and luminosity e$\gamma$ and $\gamma\gamma$
collider available, and many of the same studies as at electron-hadron
and hadron-hadron colliders can be made here.  However, one essential
difference is that, since the beam remnants (the electrons and
positrons) typically leave the detectors undetected, the energies of the
incoming photons are not known and must be reconstructed from the
properties of the final state.  In events in which most of the
final-state particles are visible in the detector, this is easily done.
However, at high energies the final-state distribution becomes
increasingly forward-peaked and much of the energy goes into the very
forward parts of the detector, or is even missed in the beam pipe.  It
is therefore essential that as much of the final state as possible is
measured, in particular to detect hadrons in the forward detectors that
have hitherto only been used for electron tagging.

It is also essential that we are able to understand the details of the
multi-particle final states in these interactions, which puts very high
demands on the event generators used in the analysis. During the course
of this workshop, some of the standard general purpose event generator
programs on the market have been developed to also handle e$\gamma$ and
$\gamma\gamma$ collisions.  This means that we can use our experience
from ep and pp collisions to give a more complete description of the
hadronic final state of two-photon collisions.  These models can now be
tested at LEP~1 and should give reliable extrapolations to LEP~2
energy.

In section~\ref{GGgeneral} of this report, we will describe briefly the
models of the generators we have studied during the workshop.  Then, in
section~\ref{GGcomp} some comparisons between the programs are presented
for different classes of events.  In section~\ref{GGmainprog}, the main
programs are presented in some detail, followed by
section~\ref{GGothers} where some other generators are presented more
briefly.  Finally in section~\ref{GGsummary} we present our conclusions.
Related work is presented in the reports of the ``$\gamma\gamma$
Physics'' \cite{GGLLggphys} and ``QCD Event Generator''
\cite{GGLLqcdgen} working groups.

\section{General Features}
\label{GGgeneral}

The event generators used for $\gamma\gamma$ physics can be divided
into two main groups.  One deals with low multiplicity final states,
like resonances, exclusive channels and leptonic channels and the
other with high mass multi-particle final states.

When there are few particles in the final state, the fully
differential cross-section for a given process can usually be derived
directly from a model of that process.  Event generators can then be
viewed as a particularly convenient numerical implementation of that
cross-section, in which arbitrarily complicated phase-space cuts and
detector simulation can be incorporated.  In this group are
`four-fermion generators', which incorporate the full set of QED
matrix elements for $\mathrm{e^+e^-\to e^+e^-f\bar{f}},$ and more
general programs that use a $\gamma\gamma$ luminosity function to
separate the process $\mathrm{e^+e^-\to e^+e^-X}$ into two stages,
$\mathrm{e^+e^-\to e^+e^-\gamma\gamma}$ and
$\gamma\gamma\to\mathrm{X}$.

The other group of generators describe multi-particle production, for
which cross-sections cannot be directly calculated in quantum
mechanics.  They use semi-classical probabilistic models to separate
the process into several phases.  First photons are radiated from the
incoming electrons to give beams of quasi-real photons.  Then a hard
sub-process is generated, using partonic $2\to2$ matrix elements folded
with the parton densities of the photon.  The emission of additional
partons from the incoming partons is generated by evolving them
``backwards'' in an initial-state parton shower, and from outgoing
partons by generating a final-state parton shower.  Finally, the
partons are converted to hadrons, which are then allowed to decay.

\subsection{Photon generation}
\label{GGflux}

A two-photon reaction
can be factorized into photon fluxes of radiation from incoming $\mathrm{e}^\pm$
and the final state of two-photon collisions.
The decomposed differential cross-section for
\begin{equation}
  \mathrm{e^+}(p_1)\mathrm{e^-}(p_2)
  \to \mathrm{e^+}(p_1')\mathrm{e^-}(p_2') \gamma(q_1)\gamma(q_2)
  \to \mathrm{e^+}(p_1')\mathrm{e^-}(p_2') \mathrm{X}(q_1+q_2)
\end{equation}
is\cite{GGSHBudnev,GGSHBonneau,GGSHField}
\begin{eqnarray}                        
  d\sigma
  &=& \frac{\alpha^2}{16\pi^4 q_1^2 q_2^2 }
    \left( \frac{(q_1 q_2)^2 -q_1^2 q_2^2 }
                {(p_1 p_2)^2 -m_1^2 m_2^2 } \right)^{1/2}
    \left( 4 \rho_1^{++}\rho_2^{++}\sigma_{TT}
          + 2|\rho_1^{+-}\rho_2^{+-}|\tau_{TT} \cos(2 \tilde{\phi})
    \right. \nonumber \\
 && \left.\;\;\;\;\;\;
          + 2\rho_1^{++}\rho_2^{00}\sigma_{TL}
          + 2\rho_1^{00}\rho_2^{++}\sigma_{LT}
          + \rho_1^{00}\rho_2^{00}\sigma_{LL}
          -8|\rho_1^{+0}\rho_2^{+0}|\tau_{TL} \cos(\tilde{\phi}) \right)
    \frac{d^3 p_1' }{E_1' }
    \frac{d^3 p_2' }{E_2' },
\label{GGSHeq:lumfun}
\end{eqnarray}
where the $\sigma$'s and $\tau$'s are linear combinations of the
cross-sections for $\gamma\gamma\rightarrow X$ of transverse(T) and
longitudinal(L) photons, the flux factor $\rho_i^{ab}$ has photon
helicities labelled by $+,-,0$.  Some dedicated $\gamma\gamma$
generators such as \GGtwogam\ \cite{GGLLtwogam} use the full form of
Eq.~\ref{GGSHeq:lumfun}, while most models simplify further by taking
the $q^2\to0$ approximation.  At $q^2 \rightarrow 0$, the photons are
quasi-real and transversely polarized and after integration over
$\tilde{\phi},$ the angle between lepton scattering planes, the only
remaining term is $\sigma_{TT}$.  Expressed in terms of a luminosity
function we have
\begin{equation}
  \sigma(e^+e^-\rightarrow e^+e^- X) = 
  \frac{d^4 {\mathcal L}_{\gamma\gamma} }
       {d\omega_1 d\omega_2 d\theta_1 d\theta_2} \sigma_{TT},
\label{GGSHeq:lumfunTT}
\end{equation}
where $w_i=E_b-E_i'$ is the photon energy in the lab frame.
The luminosity function can be decomposed as the product of a factor
for each of the photons,
\begin{equation}
  \frac{d^4 {\mathcal L}_{\gamma\gamma} }
       {d\omega_1 d\omega_2 d\theta_1 d\theta_2}
       = \frac{d^2 {\mathcal L}_{\gamma} }{d\omega_1 d\theta_1 }
         \frac{d^2 {\mathcal L}_{\gamma} }{d\omega_2 d\theta_2 },
\end{equation}
where, apart from a trivial kinematic factor, $d^2{\mathcal L}$ is the
Equivalent Photon Approximation (EPA) flux factor,
\begin{equation}
  \frac{d^2 {\mathcal L}_{\gamma} }{d\omega d\theta} =
  2p_t \; f(x,P^2),
\end{equation}
with $x$ the light-cone momentum fraction and $P^2$ the photon
virtuality, $P^2=|q^2|$.
As discussed in Ref.~\cite{GGfrixione}, there are two important
corrections to the usual EPA formula.  The first is to include the
sub-leading term of relative order $m_e^2/P^2,$
\begin{equation}
  f_{\gamma/e}(x,P^2) =
  \frac{\alpha}{2\pi}\left(\frac{1+(1-x)^2}{xP^2} -
  2m_e^2\frac{x}{P^4}\right),
\label{GGepa}
\end{equation}
which can give corrections of order 10\% for untagged and anti-tagged
cross-sections.  At present, of the QCD event generators, only
\GGphojet\ includes this correction.  The second important correction is
to include the correct, process-dependent, dynamic upper limit on
$P^2$.  However, this is only important for untagged cross-sections, and
when an anti-tag condition is imposed, as it is throughout this report,
this corrections become small.


\subsection{Photon distribution functions}

In resolved-photon processes we need parametrizations of the
distribution functions for partons inside the photon.
These obey an inhomogeneous form of the usual evolution equations.  As
discussed in more detail in the report of the ``$\gamma\gamma$ Physics''
working group\cite{GGLLggphys}, their solution can be written as the sum
of a hadronic or VMD part, which evolves according the usual homogeneous
equation, and a point-like or anomalous part.

There are a number of parametrizations available for the parton
distribution functions of on-shell photons.  Most of them are
contained in a single package, \GGpdflib\cite{GGpdflib}, to which most
of the event generators are interfaced.  At present none use the
recent models of the structure of virtual photons, although \GGherwig\
does implement a simple $P^2$-suppression model.  In the following
when comparing different generators, we use the fairly similar
SaS~1D\cite{GGRE-Schuler95a} or GRV~LO\cite{GGRE-GRV92b} sets.

\subsection{Hard sub-processes}

Having defined the structure functions and parton densities, we can
now use the same machinery as for pp or ep collisions to generate
hard sub-processes, with the exception that we here have additional
processes where the photon couples directly in the hard interaction.
In $\gamma\gamma$ collisions we therefore talk about three kinds of
events, direct, single-resolved and double resolved, depending on
whether the photons couple directly or not. In Table~\ref{GGLLtabproc}
the standard hard sub-processes are listed
for different event classes. Some programs, like \GGpythia, make a
further distinction between the anomalous and VMD--like part
of the resolved photon and therefore have six different event classes.

\begin{table}
  \begin{center}
    \begin{tabular}{|r|c|}
      \hline
      direct & $\mathrm{\gamma\gamma\to q\bar{q}}$ \\
      \hline
      single resolved & $\mathrm{\gamma q\to qg}$ \\
       & $\mathrm{\gamma g\to q\bar{q}}$ \\
      \hline
      double resolved & $\mathrm{qq'\to qq'}$ \\
       & $\mathrm{gg\to q\bar{q}}$ \\
       & \vdots \\
      \hline
       DIS & $\mathrm{eq\to eq}$ \\
      \hline
    \end{tabular}
  \end{center}
  \caption[dummy]{{\it The standard hard sub-processes used by event
                   generators for different event classes}}
  \label{GGLLtabproc}
\end{table}

In deep inelastic e$\gamma$ scattering, the exchanged photon is usually
more virtual than the struck quark, so the EPA is no longer a good
approximation (in other words the process-dependent dynamic upper limit
on $P^2$ mentioned in Section~\ref{GGflux} is exceeded).  One therefore
needs, in principle, to use the full $2\to3$ processes $\mathrm{eq \to
  eqg},$ $\mathrm{eg \to eq\bar{q}}$ and $\mathrm{e\gamma \to
  eq\bar{q}}$.  However, when the quark line is much less virtual than
the photon, it can be approximated by the DGLAP probability distribution
to find the quark inside a higher-$x$ quark, $\mathrm{q\to qg},$ gluon,
$\mathrm{g \to q\bar{q}}$ or photon, $\gamma\to\mathrm{q\bar{q}}$ and
hence can be absorbed into the evolution of the photon distribution
functions.  Thus we are again left with a $2\to2$ process, $\mathrm{eq
  \to eq},$ for which one uses the lowest-order matrix element.

Other processes that are usually treated separately are the ones
involving heavy quarks, where the matrix elements are different from
the massless case. Here there are also event generators that use the
next-to-leading order matrix elements, incorporating one additional
parton.  These can be considered as an exact treatment of the first
step of a parton shower and can be compared with the usual algorithms
discussed below, which contain approximate treatments of all steps.

\subsection{Soft processes}

The cross-section for quasi-real $\gamma\gamma$ scattering is
dominated by processes in which there is no hard scale.  Several
models exist for the subdivision of the soft cross-section into
separate components, typically: elastic, $\gamma\gamma\to\mathrm{VV}$;
single diffractive dissociation, $\gamma\gamma\to\mathrm{VX}$; double
diffractive dissociation, $\gamma\gamma\to\mathrm{XX}$; and inelastic,
$\gamma\gamma\to\mathrm{X}$.  The cross-section for each is given by
the model, and the total cross-section is simply their sum.  Clearly
the separation is model specific, for example the difference between
diffractive dissociation and inelastic scattering is the presence of a
rapidity gap between the two systems in the former and not in the
latter and the cross-sections depend on the size of this gap, and only
the sum of the processes can be directly compared between models.
Nevertheless, the available models use fairly similar definitions and
comparisons between components can prove useful.  Since most of these
reaction types cannot be calculated from first principles, they are
characterized by a rather large number of adjustable parameters.
Nevertheless, since the models assume photon-hadron duality and hadron
universality, parameters can be fixed in hadron-hadron and
photon-hadron collisions, giving parameter-free predictions for LEP~2.
\GGphojet\ and \GGpythia\ contain rather complete soft interaction models,
while \GGherwig\ contains only the non-diffractive part of the
cross-section.

\subsection{Multiple interactions}
\label{GGMHSmul}

At increasing centre-of-mass energies most sub-process cross-sections
grow faster than the total cross-section, eventually overtaking it.
Important examples include (supercritical) soft pomeron exchange and
hard two-parton scattering above any given $p_{t\mathrm{min}}$ cut.
This corresponds to the possibility of {\em multiple\/} soft or hard
scatterings within a single $\gamma\gamma$ collision. At present only
\GGphojet\ implements both soft and hard multiple scattering in the same
package with $p_{t\mathrm{min}}$ forming the boundary between the two.
\GGpythia\ and \GGherwig\ (through an interface to the \GGjimmy\ 
generator) both implement multiple hard scattering above
$p_{t\mathrm{min}},$ but with soft models that do not vary with
$p_{t\mathrm{min}},$ so that it becomes a critical parameter of the
model.  Although the general ideas of the models are similar, there are
many specific differences in their implementation.  For example, the
current models only allow multiple interactions in the hadronic
cross-section, but since the definition of anomalous and hadronic events
differs (see below), so do the multiple scattering results.  In all
three cases however, the models are essentially identical for
$\gamma\gamma$ and $\gamma$p collisions, so experience gained at HERA
will certainly help constrain the predictions for LEP~2.

\subsection{Parton showers}
\label{GGLLps}

The hard scattering disturbs the colour fields of the incoming partons
and as a result they partially `shake off' the cloud of gluons that
normally surrounds colour charges.  This gives rise to a shower of
accompanying radiation, which is conventionally modelled as a series
of emissions from the incoming parton, starting from the parton
entering the hard interaction and tracing its history back towards the
incoming photon.  During this `backwards evolution', the emission at
each step is required to be at a lower scale than the previous one
until, near the incoming photon, no more radiation is resolvable above
the infrared cutoff.  This procedure is governed by the splitting
functions of the DGLAP evolution equation and is guided by the input
set of parton distribution functions.

This is the way it is done in the two standard generators \GGherwig\ 
and \GGpythia\ where the main difference between the two is the choice
of evolution, or ordering variable -- \GGpythia\ uses the virtuality
of the evolving parton, while \GGherwig\ uses a generalized emission
angle, which incorporates colour-coherence effects.

Another important difference between the two programs is the way they
distinguish between the anomalous and VMD--like part of the structure
functions. In \GGpythia\ this distinction is done beforehand, and the
shower is identical in both cases, only using different parton densities
corresponding to the two parts of the structure functions.  In
\GGherwig, however, the two parts are treated together, and an
additional branching is introduced into the parton shower, where a quark
may be evolved back to the incoming photon. If this happens before the
shower is cut off, the event is called anomalous, if not, it is a
VMD--like event. In both cases, the photon remnant will have a larger
transverse momentum in the anomalous case --- this is automatic in the
perturbative treatment of \GGherwig, while in \GGpythia\ it is put in by
hand.

In both programs the generated partons may continue to branch in a
final-state shower, in the same way as is done in $\mathrm{e^+e^-}$
annihilation. This is discussed in more detail in the report from the
``QCD generator'' group\cite{GGLLqcdgen}.

In contrast to \GGherwig\ and \GGpythia, which use backward-evolution
algorithms, the \GGggpsone\ and \GGggpstwo\ programs evolve forwards,
i.e.\ upwards in scale toward the hard process.  This has the
advantage that the evolution of the structure function is generated by
the Monte Carlo algorithm itself, allowing a non-trivial test, whereas
backward evolution is guaranteed to reproduce whatever distribution
function is input. Like most implementations of forward evolution,
\GGggpsone\ and \GGggpstwo\ produce weighted events, which can be
inconvenient for detector simulation. Forward evolution can also be
extremely inefficient, particularly at small $x$, but this problem is
solved in \GGggpsone\ and \GGggpstwo, as described in
Section~\ref{GGGGPS12}.

In the Dipole Cascade Model, implemented in the \GGariadne\ program,
there is no explicit initial state radiation. Instead, in DIS, all
gluon emissions are described in terms of dipole radiation from the
colour dipole between the struck quark and the photon remnant. The
radiation is suppressed in the remnant direction because of its
spatial extension.

\subsection{Hadronization}

The hadronization of the produced partons is done according to the
Lund string fragmentation model \cite{GGLLstring} except in \GGherwig,
where a cluster fragmentation model is used \cite{GGLLcluster}. Both
these models are
explained in detail in the report from the ``QCD generator''
group\cite{GGLLqcdgen}.

The difference in hadronization between $\gamma\gamma$ and
$\mathrm{e^+e^-}$ events is the presence of the photon remnant in the
former. In \GGpythia\ the remnant is taken to be a single parton
(optionally quark and an antiquark in case a gluon is taken out of the
photon) which is treated like any other parton. In \GGherwig, however,
the cluster containing the remnant is treated a bit differently from the
others as described in section~\ref{GGherwig}.

The hadrons are subsequently decayed.  Those produced by the
hadronization process are generally taken to be unpolarized and thus
decay according to pure phase-space, with standard decay tables as
described in Ref.~\cite{GGLLqcdgen}.  On the other hand, when vector
mesons are produced elastically they are strongly transverse polarized,
so they are decayed accordingly, with the two pions predominantly taking
similar energies.

\section{Comparisons}
\label{GGcomp}

In the following, we will compare the predictions for LEP~2 of
different generators for different kinds of processes. Unless
specified otherwise, comparisons are made on the generator level,
using a beam energy of 87.5~GeV, a total integrated luminosity of
500~$\mathrm{pb}^{-1}$ and requiring at least an invariant mass of 2~GeV
of particles in the region $|\cos\theta|<0.97$.

\subsection{Exclusive channels and resonances}

\begin{figure}
  \centering\epsfig{file=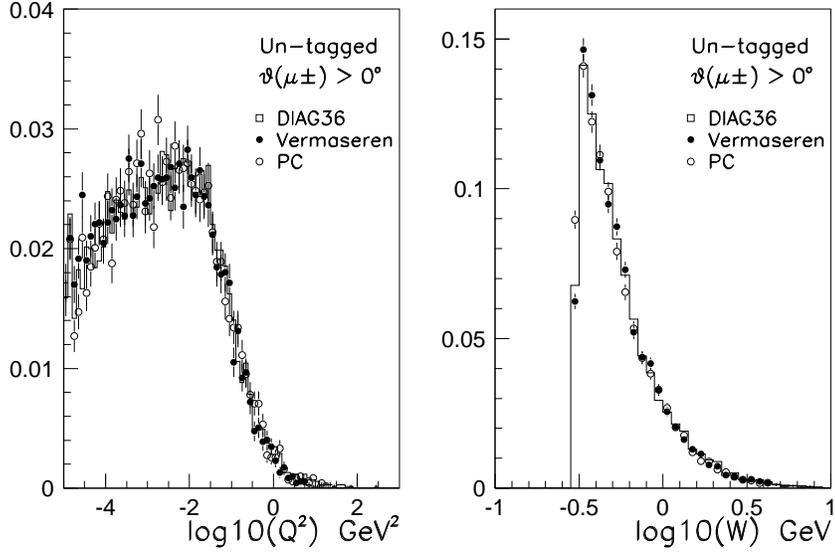, width=.75\linewidth}
  \caption{{\it The $Q^2$ and $W$ distributions
      of $e^+e^-\rightarrow e^+e^-\mu^+\mu^-$ at
      $\protect\sqrt{s}=m_{\scriptscriptstyle Z}$.}}
  \label{GGSHfig:LumiQED}
\end{figure}

Lepton pair production, $\mathrm{e^+e^-\to e^+e^-\ell^+\ell^-}$
is well described by the exact matrix elements, which are
dominated by the multiperipheral diagram.
In terms of the luminosity function, the lepton pair production 
is given by the QED structure function for 
$\gamma\gamma \to \mathrm{\ell^+\ell^-}$.
We have compared muon pair production 
by \GGpc\cite{GGSHPC} applying the luminosity function
to the exact calculations of matrix elements
by \GGdiag36\cite{GGSHDIAG36} and Vermaseren\cite{GGSHVermaseren}.
The mass threshold imposed on the pair is $W>300$ MeV. 
The distributions of $Q^2=\max(-q_1^2,-q_2^2)$ and $W$ obtained are shown
in Fig.~\ref{GGSHfig:LumiQED}.

The cross-section of $\gamma\gamma\rightarrow R$ for a narrow resonance
of spin-$J$ is given by\cite{GGSHBudnev,GGSHPoppe}
\begin{equation}
  \sigma(\gamma\gamma\rightarrow R) =
  8 \pi (2J+1) F^2(q_1^2,q_2^2)
  \frac{\Gamma_{\gamma\gamma} \Gamma }
       {(m_R^2-W^2)^2 +m_R^2 \Gamma^2 }.
\label{GGSHeq:BW}
\end{equation}
Here the resonance has mass $m_R$,
radiative width $\Gamma_{\gamma\gamma}$,
and energy dependent width $\Gamma$.
For a wide resonance the Breit-Wigner term of Eq.~\ref{GGSHeq:BW}
is multiplied by $(m_R/W)^n$ \cite{GGSHPoppe}, with
$n$=1,2 for the resonance decaying into two, three 
stable particles respectively.
The $q^2$ dependence is given by the form factor 
function that satisfies $F^2(0,0)=1$. The VMD model predicts
\begin{equation}
  F^2(q_1^2,q_2^2)= \sum_{V_1,V_2} \frac{A_{V_1}}{(1-q_1^2/m_{V_1}^2)^2} \;
  \frac{A_{V_2}}{(1-q_2^2/m_{V_2}^2)^2},
  \hspace{1.cm} \sum_V A_V\equiv1,
  \hspace{1.cm} V=\rho,\omega,\phi,J/\psi \ldots
\end{equation}

In generating two-photon resonances, the mass spectrum is the product of
the Breit-Wigner distribution and the two-photon invariant mass spectrum
obtained from the luminosity function.
The decay products are further described by the
matrix element according to the spin-parity and helicity
state that couple to the angular distributions
of the final state particles. 

\begin{figure}
  \vspace{-5mm}
  \centering\epsfig{file=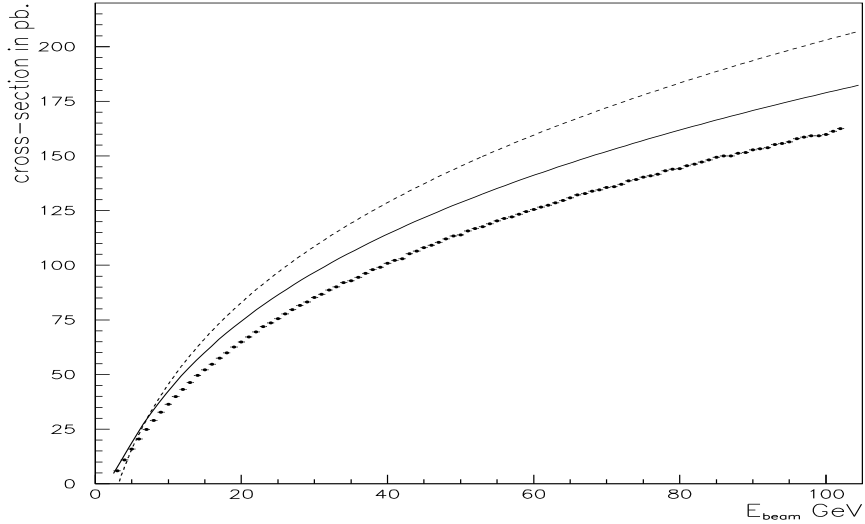,height=8.cm,width=.75\linewidth}
  \vspace{-5mm}
  \caption{{\it The $\eta_c$ cross-section versus beam energy: Low
    approximation, dashed; narrow-width approximation, solid; \GGrespro\
    results, points.}}
  \label{GGSHfig:EtacXsection}
\end{figure}

We have compared the cross-section generated for $\eta_c$ resonance as a
function of beam energy which is shown in
Fig.~\ref{GGSHfig:EtacXsection}.  The dashed line is the Low
approximation~\cite{GGSHLow} (i.e.\ the narrow-width approximation and
the leading term of the EPA, but with the $x$-dependence of the $P^2$
integration neglected: $\int dP^2/P^2 \to \log s/m_e^2$), the solid line
is the luminosity function of Eq.~\ref{GGSHeq:lumfun} using the
approximation of $W^2=4 w_1 w_2$ for small $q^2$, and the points are the
\GGrespro\ \cite{GGSHRESPRO} calculation using exact $w_1 w_2$ without
form factor.

\begin{figure}[t]
  \vspace{-5mm}
  \centering\epsfig{file=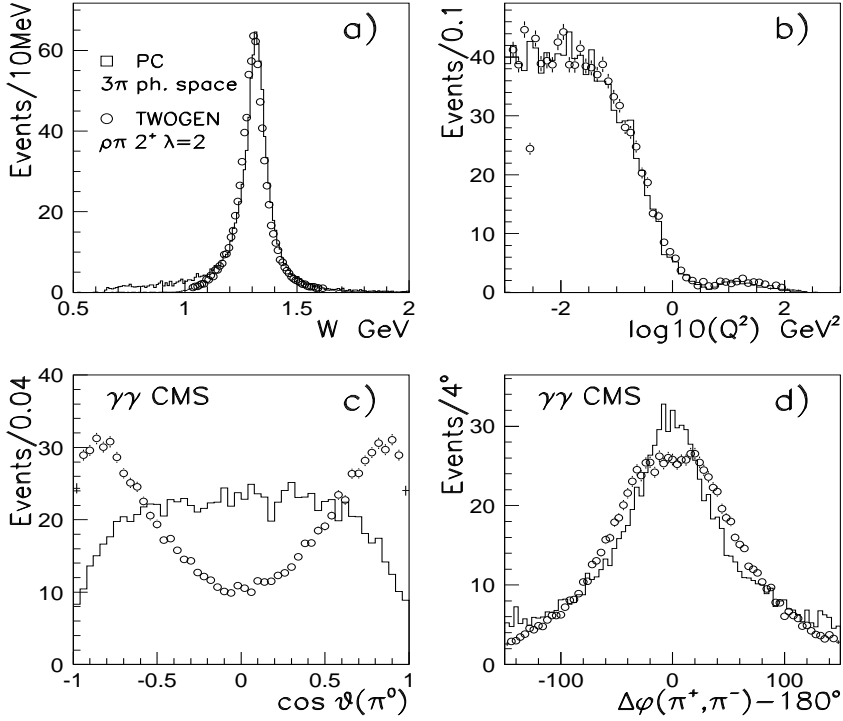,height=11cm,width=.75\linewidth}
  \vspace{-5mm}
  \caption{{\it The $a_2(1320)$ simulations of \GGpc\ in 3$\pi$ phase
            space and \GGtwogen\ with $J^P$=$2^+,$
            helicity $\lambda$=2 and intermediate
            state $\rho\pi$. Detector acceptance are set
            to $\theta > 10^\circ,25^\circ$ for calorimeter and
            central tracker respectively.}}
  \label{GGSHfig:a2com}
\end{figure}

We have instrumented the photon flux generated by \GGpc\ and
\GGtwogen\cite{GGLLtwogen} to simulate $a_2(1320)$ mesons in
$\pi^+\pi^-\pi^0$ final states at LEP~1 with detector acceptance set
to $\theta > 10^\circ,25^\circ$ from the beam direction for
calorimeter and central tracker respectively.  Shown in
Fig.~\ref{GGSHfig:a2com}a and b are the invariant mass and $Q^2$
distributions of $a_2(1320)$ generated by \GGpc\ and \GGtwogen.
Instrumented in \GGpc\ is a narrow resonance of Eq.~\ref{GGSHeq:BW}
decaying into $3\pi$ phase space, while in \GGtwogen\ the Breit-Wigner
of a wide resonance is multiplied by a matrix element of $J^P$=$2^+$
with helicity $\lambda$=2 and intermediate state of $a_2\rightarrow
\rho\pi \rightarrow \pi^+\pi^-\pi^0$.  A second Breit-Wigner term for
$\rho$ suppresses events at low $W$.  Within the detector acceptance
good agreement is seen.  The coupling to spin-parity and helicity
state is demonstrated in Fig.~\ref{GGSHfig:a2com}c and d in which the
$\pi^0$ distribution in polar angle and acollinear azimuthal angle of
the two charged pion tracks are compared with the phase space
distributions of \GGpc.

\subsection{Minimum bias events}
\label{GGminb}

\input LEP2-GGEG-MB

\subsection{Deep inelastic scattering}
\label{GGMHSdis}

When measuring the photon structure function in deep inelastic e$\gamma$
scattering, it is of course necessary to be able to accurately measure
the $Q^2$ and the Bjorken-$x$ in each event. The $Q^2$ is easily
obtained from the energy and angle of the scattered electron, but since
the energy of the target photon is unknown, $x$ must be determined from
the hadronic final state.  Thus one measures the distribution of visible
hadronic mass, $W_{\mathrm{vis}},$ and hence $x_{\mathrm{vis}}\equiv
Q^2/(W_{\mathrm{vis}}^2+Q^2)$.  This is then converted into the
distribution of $x$ using an unfolding procedure, typically based on
Refs.~\cite{GGLLblobel} or~\cite{GGLLbayes}.

It is clear that the more of the solid angle over which we detect
hadrons, the better correlated $W_{\mathrm{vis}}$ will be with the true
hadronic mass, $W_{\mathrm{true}}$, so the less work the unfolding
procedure will have to do.  For now we simply define $W_{\mathrm{vis}}$
to be the total invariant mass of all the hadrons within the angle
covered by the tracking system of a typical LEP detector,
$|\cos\theta|<0.97,$ as done in most previous analyses, and return to
this point later.

The unfolding procedure relies heavily on the event generator's ability
to correctly model the hadronic final state.  In previous analyses (see
for example Refs.~\cite{GGLLopal,GGLLdelphi}), the consistency of the
generators used for the unfolding has been checked using inclusive
distributions of the data, such as the distributions in $Q^2,$
$W_{\mathrm{vis}}$ and particle multiplicity and momentum distributions.
The problem is that such distributions can be fitted with basically any
generator by adjusting the number of events in each $x$ and $Q^2$ bin,
i.e.\ by modifying the input parton distribution.  To really check the
generator's description of the final states, one needs to investigate
less inclusive distributions.

Experience from HERA shows that at small $x,$ the distribution of
transverse energy, $E_\perp,$ in the proton direction is very difficult
to describe.  Indeed, programs with conventional ``DGLAP-like''
initial-state parton showers cannot explain the measured $E_\perp$
perturbatively and are dominated by the non-perturbative components.  On
the other hand, the dipole cascade in the \GGariadne\ program describes
the $E_\perp$ flow rather well at the perturbative level, with only
small hadronization corrections.  While more detailed measurements might
hope to distinguish them, the most recent versions of all models are in
good agreement with current data.

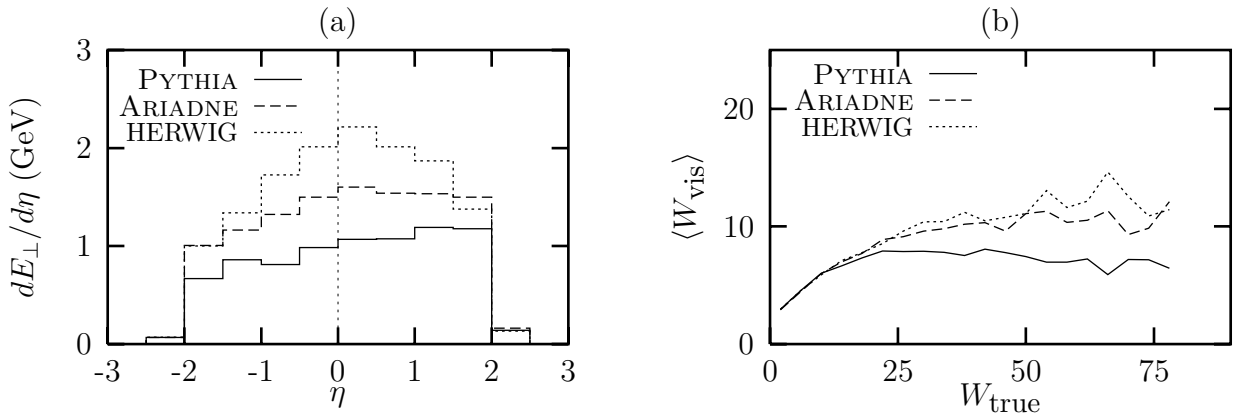
\begin{figure}
\setlength{\unitlength}{0.1bp}
\special{!
/gnudict 40 dict def
gnudict begin
/Color false def
/Solid false def
/gnulinewidth 5.000 def
/vshift -33 def
/dl {10 mul} def
/hpt 31.5 def
/vpt 31.5 def
/M {moveto} bind def
/L {lineto} bind def
/R {rmoveto} bind def
/V {rlineto} bind def
/vpt2 vpt 2 mul def
/hpt2 hpt 2 mul def
/Lshow { currentpoint stroke M
  0 vshift R show } def
/Rshow { currentpoint stroke M
  dup stringwidth pop neg vshift R show } def
/Cshow { currentpoint stroke M
  dup stringwidth pop -2 div vshift R show } def
/DL { Color {setrgbcolor Solid {pop []} if 0 setdash }
 {pop pop pop Solid {pop []} if 0 setdash} ifelse } def
/BL { stroke gnulinewidth 2 mul setlinewidth } def
/AL { stroke gnulinewidth 2 div setlinewidth } def
/PL { stroke gnulinewidth setlinewidth } def
/LTb { BL [] 0 0 0 DL } def
/LTa { AL [1 dl 2 dl] 0 setdash 0 0 0 setrgbcolor } def
/LT0 { PL [] 0 1 0 DL } def
/LT1 { PL [4 dl 2 dl] 0 0 1 DL } def
/LT2 { PL [2 dl 3 dl] 1 0 0 DL } def
/LT3 { PL [1 dl 1.5 dl] 1 0 1 DL } def
/LT4 { PL [5 dl 2 dl 1 dl 2 dl] 0 1 1 DL } def
/LT5 { PL [4 dl 3 dl 1 dl 3 dl] 1 1 0 DL } def
/LT6 { PL [2 dl 2 dl 2 dl 4 dl] 0 0 0 DL } def
/LT7 { PL [2 dl 2 dl 2 dl 2 dl 2 dl 4 dl] 1 0.3 0 DL } def
/LT8 { PL [2 dl 2 dl 2 dl 2 dl 2 dl 2 dl 2 dl 4 dl] 0.5 0.5 0.5 DL } def
/P { stroke [] 0 setdash
  currentlinewidth 2 div sub M
  0 currentlinewidth V stroke } def
/D { stroke [] 0 setdash 2 copy vpt add M
  hpt neg vpt neg V hpt vpt neg V
  hpt vpt V hpt neg vpt V closepath stroke
  P } def
/A { stroke [] 0 setdash vpt sub M 0 vpt2 V
  currentpoint stroke M
  hpt neg vpt neg R hpt2 0 V stroke
  } def
/B { stroke [] 0 setdash 2 copy exch hpt sub exch vpt add M
  0 vpt2 neg V hpt2 0 V 0 vpt2 V
  hpt2 neg 0 V closepath stroke
  P } def
/C { stroke [] 0 setdash exch hpt sub exch vpt add M
  hpt2 vpt2 neg V currentpoint stroke M
  hpt2 neg 0 R hpt2 vpt2 V stroke } def
/T { stroke [] 0 setdash 2 copy vpt 1.12 mul add M
  hpt neg vpt -1.62 mul V
  hpt 2 mul 0 V
  hpt neg vpt 1.62 mul V closepath stroke
  P  } def
/S { 2 copy A C} def
end
}
\begin{picture}(2519,1511)(100,0)
\special{"
gnudict begin
gsave
50 50 translate
0.100 0.100 scale
0 setgray
/Helvetica findfont 100 scalefont setfont
newpath
-500.000000 -500.000000 translate
LTa
600 251 M
1736 0 V
-868 0 R
0 1109 V
LTb
600 251 M
63 0 V
1673 0 R
-63 0 V
600 621 M
63 0 V
1673 0 R
-63 0 V
600 990 M
63 0 V
1673 0 R
-63 0 V
600 1360 M
63 0 V
1673 0 R
-63 0 V
600 251 M
0 63 V
0 1046 R
0 -63 V
889 251 M
0 63 V
0 1046 R
0 -63 V
1179 251 M
0 63 V
0 1046 R
0 -63 V
1468 251 M
0 63 V
0 1046 R
0 -63 V
1757 251 M
0 63 V
0 1046 R
0 -63 V
2047 251 M
0 63 V
0 1046 R
0 -63 V
2336 251 M
0 63 V
0 1046 R
0 -63 V
600 251 M
1736 0 V
0 1109 V
-1736 0 V
600 251 L
LT0
1152 1249 M
180 0 V
600 251 M
145 0 V
0 25 V
144 0 V
0 222 V
145 0 V
0 71 V
145 0 V
0 -18 V
144 0 V
0 64 V
145 0 V
0 31 V
145 0 V
0 2 V
144 0 V
0 43 V
145 0 V
0 -5 V
145 0 V
0 -383 V
144 0 V
0 -52 V
145 0 V
LT1
1152 1149 M
180 0 V
600 251 M
145 0 V
0 25 V
144 0 V
0 347 V
145 0 V
0 58 V
145 0 V
0 59 V
144 0 V
0 65 V
145 0 V
0 38 V
145 0 V
0 -23 V
144 0 V
0 -2 V
145 0 V
0 -13 V
145 0 V
0 -494 V
144 0 V
0 -60 V
145 0 V
LT3
1152 1049 M
180 0 V
600 251 M
145 0 V
0 26 V
144 0 V
0 344 V
145 0 V
0 125 V
145 0 V
0 143 V
144 0 V
0 106 V
145 0 V
0 75 V
145 0 V
0 -75 V
144 0 V
0 -53 V
145 0 V
0 -182 V
145 0 V
0 -460 V
144 0 V
0 -49 V
145 0 V
stroke
grestore
end
showpage
}
\put(1092,1049){\makebox(0,0)[r]{{\footnotesize HERWIG}}}
\put(1092,1149){\makebox(0,0)[r]{{\small \GGariadne}}}
\put(1092,1249){\makebox(0,0)[r]{{\small \GGpythia}}}
\put(1468,1460){\makebox(0,0){(a)}}
\put(1468,51){\makebox(0,0){$\eta$}}
\put(350,805){%
\special{ps: gsave currentpoint currentpoint translate
270 rotate neg exch neg exch translate}%
\makebox(0,0)[b]{\shortstack{$dE_\perp/d\eta$ (GeV)}}%
\special{ps: currentpoint grestore moveto}%
}
\put(2336,151){\makebox(0,0){3}}
\put(2047,151){\makebox(0,0){2}}
\put(1757,151){\makebox(0,0){1}}
\put(1468,151){\makebox(0,0){0}}
\put(1179,151){\makebox(0,0){-1}}
\put(889,151){\makebox(0,0){-2}}
\put(600,151){\makebox(0,0){-3}}
\put(540,1360){\makebox(0,0)[r]{3}}
\put(540,990){\makebox(0,0)[r]{2}}
\put(540,621){\makebox(0,0)[r]{1}}
\put(540,251){\makebox(0,0)[r]{0}}
\end{picture}
\input GGDIS2.tex
\caption[dummy]{{\it (a) The transverse energy flow in deep inelastic
    e$\gamma$ scattering at LEP~2 as predicted by different generators
    for $x<0.01$. (b) The mean value of $W_{\mathrm{vis}}$ as a
    function of $W_{\mathrm{true}}$ at LEP~2 as predicted by
    different generators.}}
\label{GGLLetflow}
\end{figure}

\begin{figure}[t]
\begin{center}
\epsfig{file=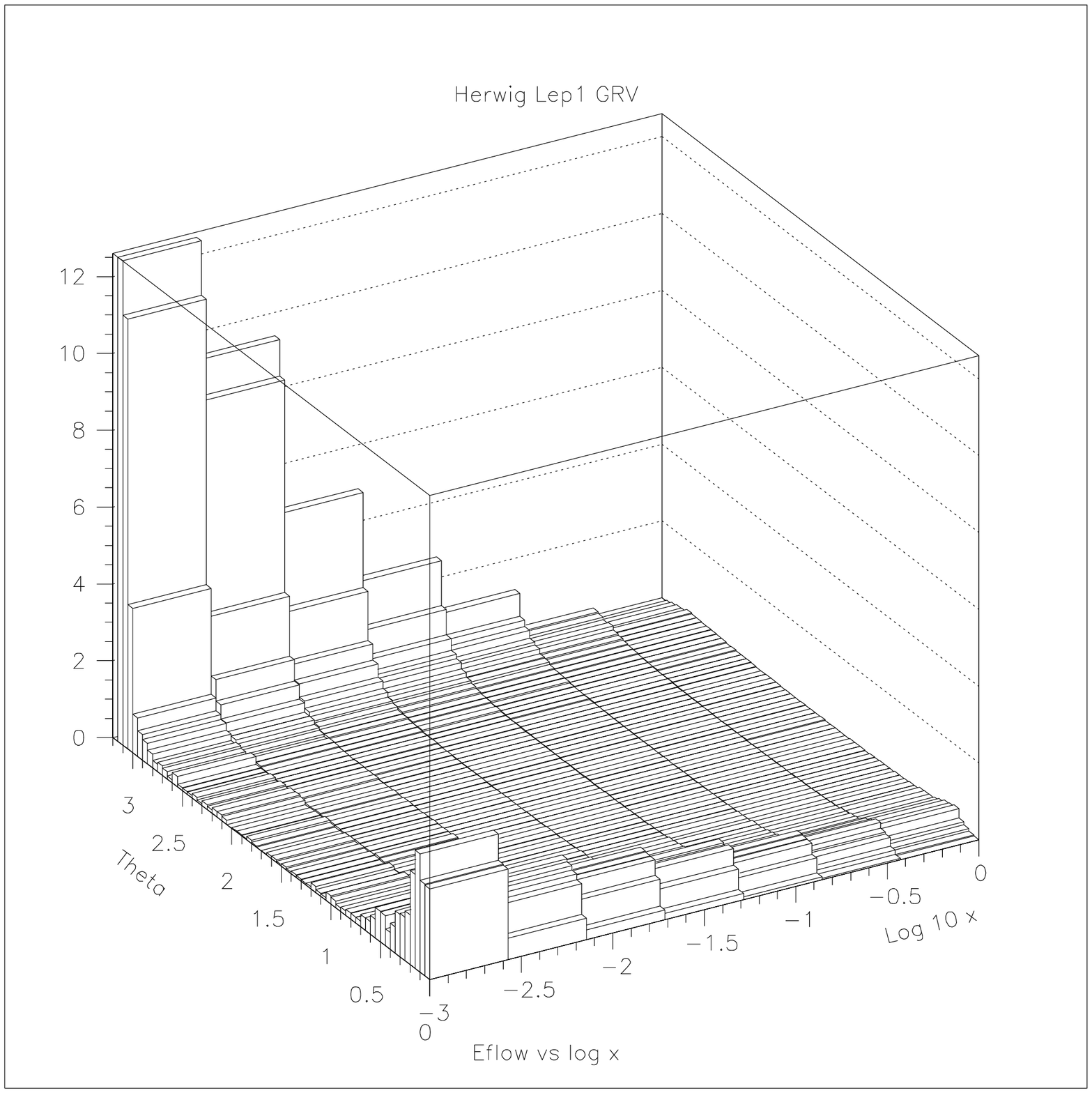,width=1.0\textwidth,height=12cm}  
\vspace*{-10mm}
\end{center}
\caption[dummy]{{\it The angular distribution of energy flow as a function
    of $x$ at LEP~1 as predicted by \GGherwig\ for events with
    $4<Q^2<30$ GeV$^2$. The axes are such that the tagged electron is
    always on the forward, $0^\circ$ side, while the photon remnant
    goes in the backward direction.}}
\label{GGLLmckigney}
\end{figure}

In Fig.~\ref{GGLLetflow}a we show the $E_\perp$ flow for small-$x$
events at LEP~2 predicted by some event generators. The reach in $x$
is not as large as at HERA, but the differences are still large. In
the forward and central regions, the differences are mostly due to the
fact that \GGherwig\ corrects the hardest emission to reproduce the
full ${\cal O}(\alpha^2\alpha_S)$ and ${\cal O}(\alpha^3)$ matrix
elements, while this is only done partly in \GGariadne\ and not at all
in \GGpythia. In the backward (photon remnant) direction, \GGpythia\ 
is expected to be lower than \GGariadne\ because of the differences in
the parton showers as explained above. \GGherwig\ is here higher than
\GGpythia\ due to the special treatment of the remnant fragmentation.

It is clear that the relationship between $W_{\mathrm{vis}}$ and
$W_{\mathrm{true}}$ is closely related to the energy flow, and in
Fig.~\ref{GGLLetflow}b we see that the models that give higher $E_\perp$
flow also give a stronger correlation between $W_{\mathrm{vis}}$ and
$W_{\mathrm{true}}$ as expected.  Also, all models predict a very weak
correlation at large $W_{\mathrm{true}},$ because more and more of the
photon remnant falls outside the assumed acceptance as $x$ gets smaller.
This is demonstrated more clearly in Fig.~\ref{GGLLmckigney}, where the
angular distribution of energy flow as a function of $x$ is shown for
LEP~1.  The situation is even more extreme at LEP~2.  The amount of
energy falling into the central regions of the detector hardly increases
with $W,$ while almost all the increase is concentrated in the far
forward and backward regions.

Our assumed detector acceptance covers all but the last five bins at
each end, but we see that this is where most of the energy increase
lies.  However, all the LEP experiments have detectors in this region,
which are used to tag electrons, and cover all but the very last bin of
this plot.  Thus if these could be used, even just to sample some of the
hadronic energy in this region, a greatly improved $W$ measurement could
be made.  Preliminary studies indicate that in addition to the neutral
pions, which can be measured because they decay to photons, a
significant fraction of charged pions deposit some of their energy on
the way through the detector.  As a generator-level prescription to get
some measure of how big an improvement this will make, we assume that
neutral pions are perfectly measured, while no other hadrons are
measured at all.  In addition, we multiply all $\pi^0$ energies by a
factor of three to arrive at an estimate of the total hadronic energy in
the forward region.

In addition to extending our angular coverage, we can use additional
kinematic variables of the final state to improve our $W$ measurement.
Defining the light-cone components of the particle momenta $p_\pm\equiv
E\pm p_z$ we can write the invariant mass of the hadronic system as
\begin{equation}
  W^2 = \biggl(\sum_i p_{i+}\biggr)\biggl(\sum_i p_{i-}\biggr) -
  \biggl|\sum_i \vec{p}_{i\perp}\biggr|^2,
  \label{GGLLlight}
\end{equation}
where $i$ runs over all hadrons.  Using energy-momentum conservation
(and neglecting the virtuality of the target photon) we can get some of
the terms in Eq.~\ref{GGLLlight} from the scattered electron, to define
\begin{equation}
  W_{\mathrm{rec}}^2 \equiv \biggl(p_{e+}-p_{e'+}\biggr)
  \biggl(\sum_i p_{i-}\biggr) - \biggl|\vec{p}_{e'\perp}\biggr|^2.
  \label{GGLLlightagain}
\end{equation}
This is equivalent to the well-known Jacquet-Blondel idea in
photoproduction\cite{GGJacqBlon} except for the inclusion of the
transverse momentum component.  The sum over $i$ now runs over hadrons
in the central region, $|\cos\theta|<0.97,$ and $\pi^0$s in the forward
region, $0.97 < |\cos\theta| < 0.9996,$ and we include an extra factor
of three for the forward $\pi^0$s.

Using Eq.~\ref{GGLLlightagain} has two advantages over the na\"\i ve
method of just using the invariant mass of detected hadrons.  Firstly,
the detector resolution enters as the product of hadronic and leptonic
resolutions, rather than as hadronic-squared which, since the leptonic
resolution is usually much better than the hadronic, gives improved
overall resolution.  Secondly, the effect of missing particles in the
current direction $\theta\sim0$ is minimized, because they give a
negligible contribution to $E-p_z$.  This, in addition to the inclusion
of forward $\pi^0$s, means that $W_{\mathrm{rec}}$ is much better
correlated with the true $W,$ as seen in Fig.~\ref{GGLLwwrec}a.  Also
the differences between the models is much smaller.

\begin{figure}
\input GGDIS4.tex
\input GGDIS3.tex
\caption[dummy]{{\it (a) The mean value of $W_{\mathrm{rec}}$ as a function
    of $W_{\mathrm{true}}$ at LEP~2 as predicted by different
    generators. (b) The mean value of $W_{\mathrm{vis}}$ for
    anomalous and normal events as a function of $W_{\mathrm{true}}$ at
    LEP~2 as predicted by different generators.}}
\label{GGLLwwrec}
\end{figure}

\begin{figure}
\begin{center}
\epsfig{file=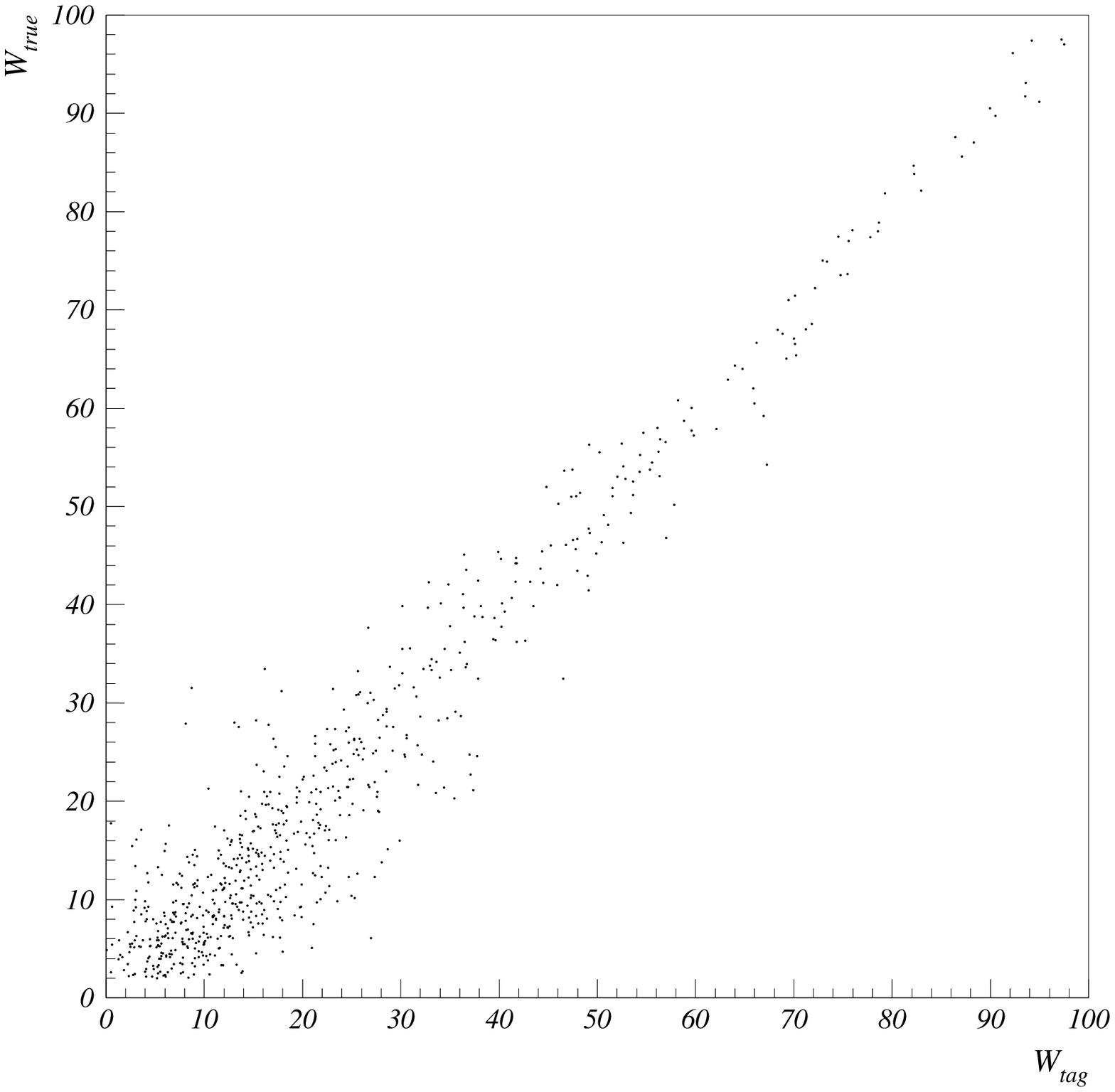,width=1.0\textwidth,height=8.5cm}  
\vspace*{-10mm}
\end{center}
\caption[dummy]{{\it Scatterplot of $W_{\mathrm{true}}$ vs.\ 
    $W_{\mathrm{tag}}$ at LEP~2 as predicted by \GGherwig\ for
    double-tag events after detector simulation.}}
\label{GGLLalison}
\end{figure}

In anomalous events, the photon remnant typically has larger
transverse momentum, which means that in such events the correlation
between $W_{\mathrm{vis}}$ and $W_{\mathrm{true}}$ is
higher, as is seen in Fig.~\ref{GGLLwwrec}b. The definition of
anomalous is, however, not the same in all generators, and in
\GGherwig, where anomalous events are defined by having a large
transverse momentum remnant, the differences are larger than in
\GGpythia, where the remnant has larger transverse momentum only on
average.

Finally in double-tag events, we have more accurate information about
the target photon energy, and hence the $W_{\gamma\gamma},$ as is seen
in Fig.~\ref{GGLLalison}. It may be possible to use such events to
calibrate the unfolding procedure for single-tag events. However, care
must be taken, as in double-tag events, the target photon is more
off-shell and the fraction of anomalous events are expected to be
higher. See also sections 2 and 5 of the report from the
``$\gamma\gamma$ Physics'' \cite{GGLLggphys} working group for more
discussions of double-tag events.

\subsection{\boldmath High-$p_\perp$}

Next-to-leading order QCD predictions are now available for both hadron
and jet distributions in $\gamma\gamma$ collisions.  Comparisons of data
with these predictions are hoped to provide additional constraints on
the parton content of the photon, particularly the gluon.  However they
apply to the partonic final state, consisting of at most three quarks or
gluons, and cannot be compared to data without incorporating
hadronization effects.  For this to be done meaningfully, it is
important to use general purpose QCD event generators, so that the
parton level is as accurate a representation of the calculation as
possible, while the hadronization is well-constrained by other
reactions.  Indeed the problems faced in $\gamma\gamma$ are almost
identical to those in $\gamma$p photoproduction at HERA, and we can
expect that they will have been explored in detail by the time LEP2
analysis starts.

The hadronization effects can be broadly split into two groups:
hadronization itself, and `underlying event' effects, the latter
coming principally from twice-resolved events in which the two photon
remnants interact with each other in addition to the main scattering.
Of course this separation is model-dependent, as hadrons described as
coming from initial-state radiation in one model could be described as
an underlying event in another, but it is a useful guide to where we
expect the models to be more or less reliable.  Hadronization itself
is expected to be largely process-independent so models tuned to
$\mathrm{e^+e^-}$ annihilation should give a reasonable description of
data.  Underlying event effects are much more poorly understood and
the models are almost completely unconstrained at present.  Their main
effect is to spray additional transverse energy around the event,
which can severely distort jet measurements, principally by adding
extra transverse momentum to the jets.  Because the jet spectrum is so
rapidly falling, this increases the jet cross-section considerably at
any given jet transverse momentum.

\begin{figure}
  \centerline{\epsfig{figure=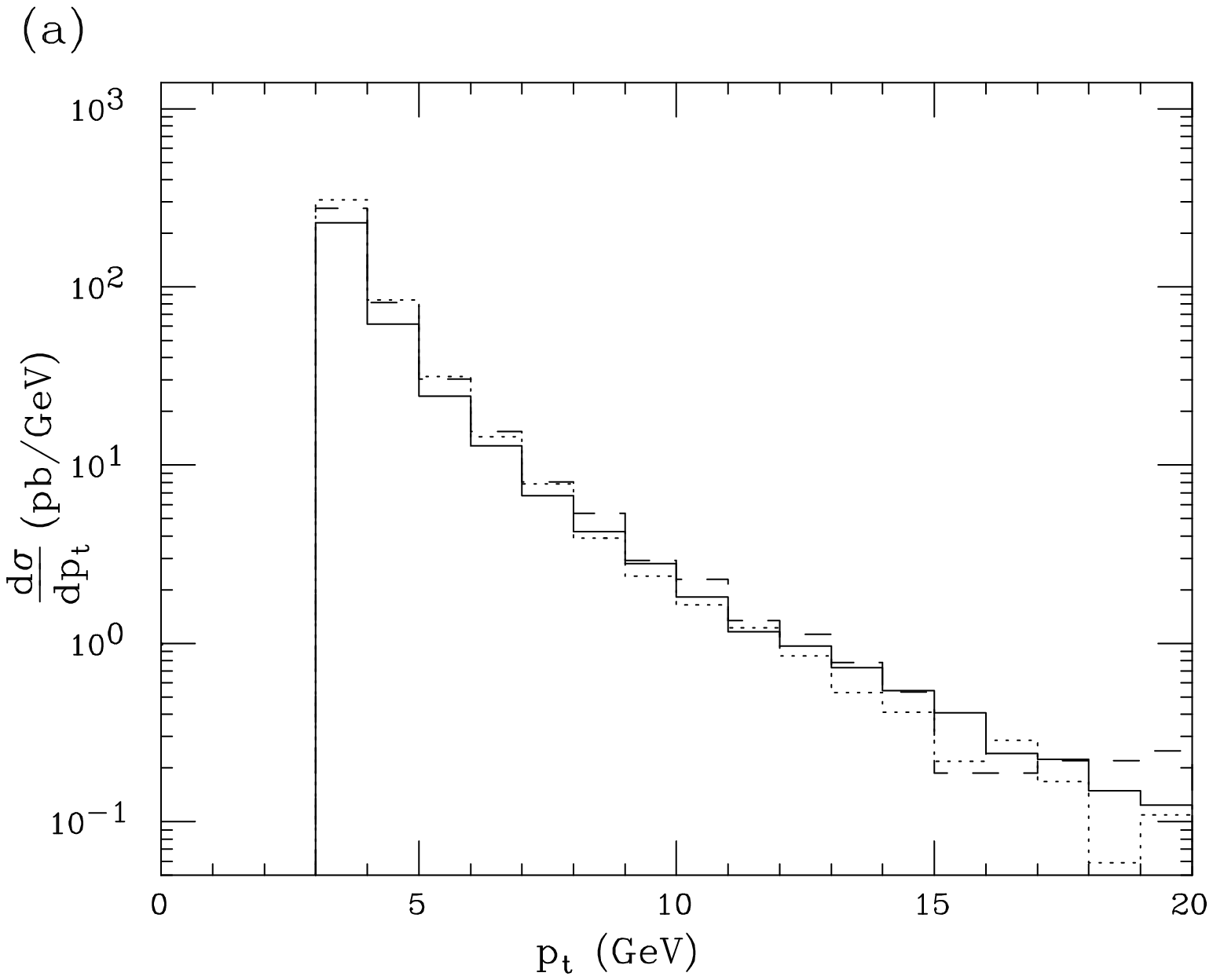,height=6cm}\hspace{\fill}
              \epsfig{figure=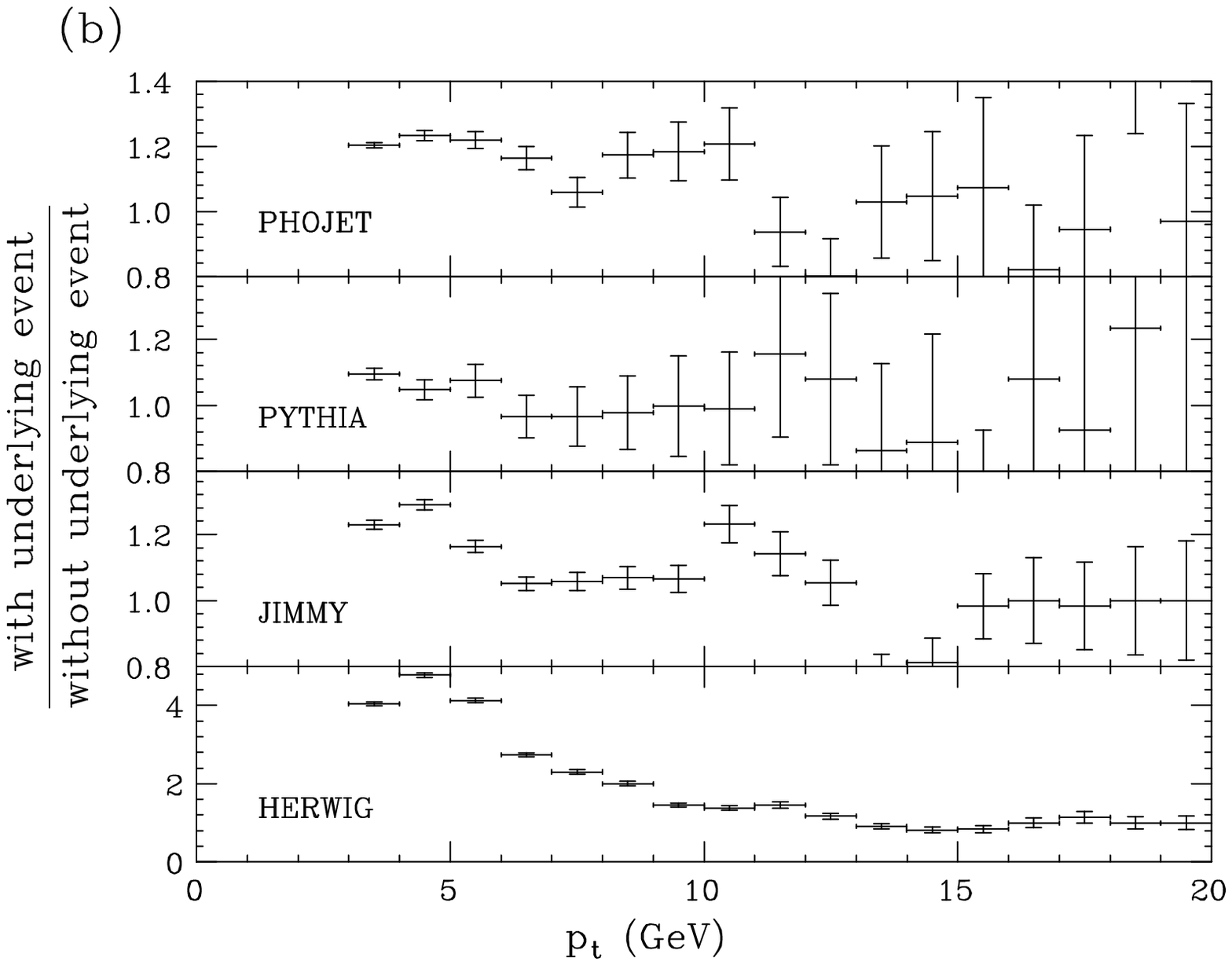,height=6cm}}
  \caption[]{\it{The inclusive jet cross-section as a function of $p_t$
      according to \GGherwig\ (solid), \GGpythia\ (dashed) and
      \GGphojet\ (dotted), when all their underlying event models are
      turned off~(a), and the relative changes when they are turned
      on~(b).  Errors shown come purely from the statistics of the
      Monte Carlo samples.}}
  \label{GGMHSpt}
\end{figure}
In Fig.~\ref{GGMHSpt}, we show the inclusive jet cross-section, as a
function of the jet transverse momentum.  Jets are reconstructed using
the CDF cone algorithm with a cone size of 1.0 and a minimum transverse
momentum of 3~GeV, using all hadrons within the angular acceptance,
$|\cos\theta|<0.97$.  We see reasonable agreement between \GGpythia,
\GGphojet\ and \GGherwig\ when their underlying event and multiple
scattering models are turned off.  However, while all of them predict a
significant increase with multiple scattering and the underlying event,
there is little agreement about its size.  It is worth pointing out that
\GGherwig's soft underlying event model produces far too big an effect
to fit HERA photoproduction data, one of the motivations for
incorporating the \GGjimmy\ multiple hard interaction model into
\GGherwig\ instead.  Amongst the three other models the relative effect
of multiple interactions is comparable to the differences between the
models.  It is clear that these effects must be understood before an
accurate jet measurement can be made below about 10 GeV.

Of course the data itself can be used to study the effects and constrain
the models.  The jet energy profile is particularly sensitive to the
underlying event, since perturbative radiation and hadronization are
concentrated at the core of jets, while the underlying background is
much more diffuse.  Furthermore, since one expects the underlying event
to mainly be important in the twice-resolved process, if we make a
physical separation of direct and resolved photons we can test this
picture.  At HERA, a cut is made on $x_\gamma,$ the fraction of the
photon's light-cone momentum carried by the reconstructed
jets\cite{GGzeusdijets}.  In fact the HERA experiments define this in
dijet events from the two hardest jets, while we propose a slightly
different approach for $\gamma\gamma$ collisions: to use \emph{all\/}
the reconstructed jets, regardless of how many there are.  Clearly when
this fraction is close to 1 there can be no photon remnant, whereas when
it is much smaller than 1 there must be one.  The cut is typically set
at around 0.7.  For $\gamma\gamma$ collisions, we define a \emph{pair\/}
of momentum fractions by analogy, as
\begin{equation}
  x_\gamma^\pm = \frac
  {\sum_{\mathrm{jets}}(E \pm p_z)}{\sum_{\mathrm{particles}}(E \pm p_z)},
\label{GGMHSxgtpm}
\end{equation}
where, following the discussion in section~\ref{GGMHSdis}, we include
three times the forward $\pi^0$ energy in the sum over particles, which
significantly improves the measurement of the denominator of
Eq.~\ref{GGMHSxgtpm}.  We define the axes such that
$x_\gamma^+>x_\gamma^-$.

\begin{figure}
  \centerline{\epsfig{figure=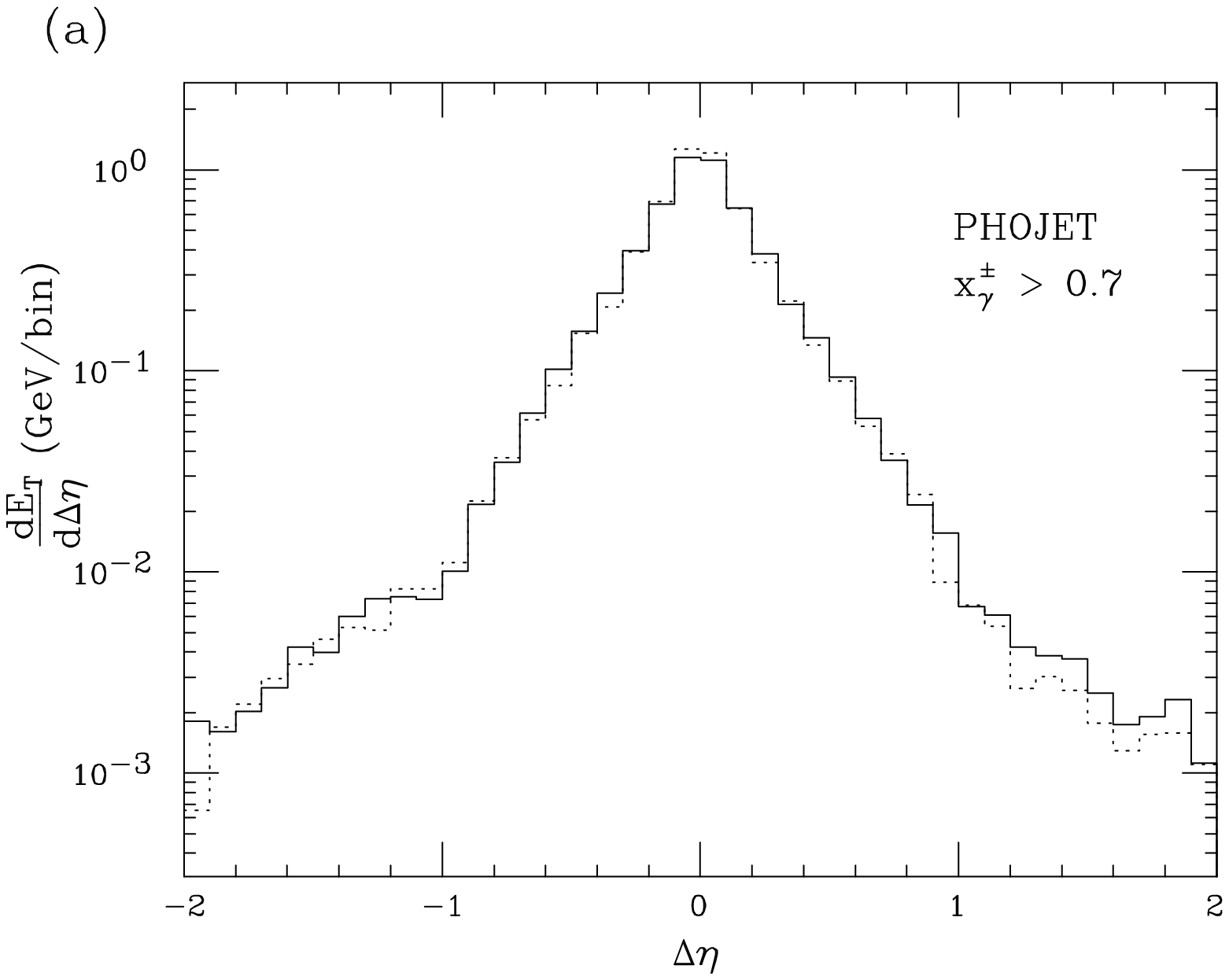,height=6cm}\hspace{\fill}
              \epsfig{figure=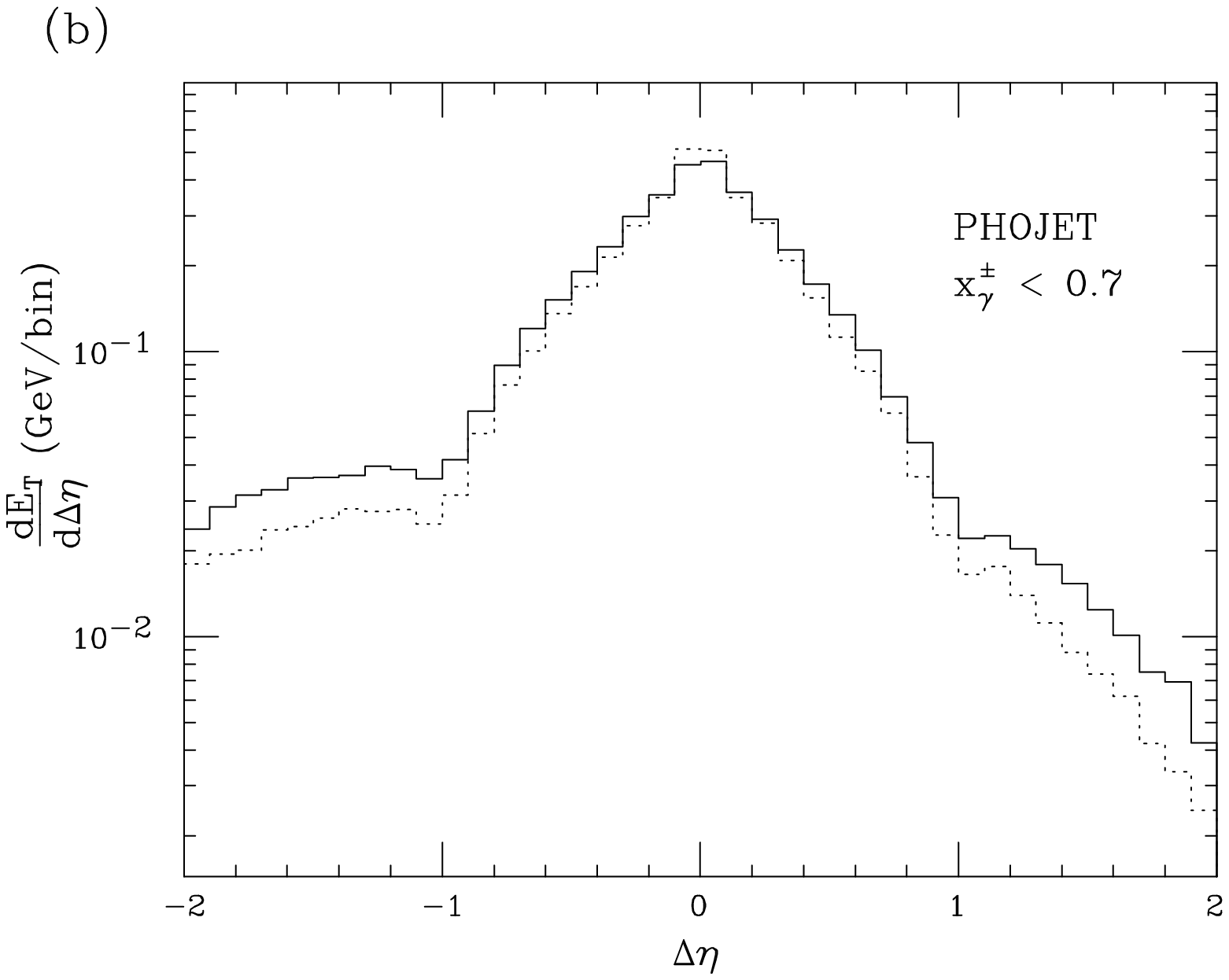,height=6cm}}
  \caption[]{\it{The jet profile of central jets divided into (a) direct
      events and (b) double-resolved events according to a physical
      definition, as predicted by \GGphojet\ with (solid) and without
      (dotted) multiple interactions.  $\Delta\eta$ is the rapidity
      relative to the jet's, with the axes defined such that
      $x_\gamma^+>x_\gamma^-$ such that there is always more of a
      remnant to the left than the right.}}
  \label{GGMHSjet}
\end{figure}
In Fig.~\ref{GGMHSjet}, we show the $E_T$ profile of the hardest jet in
the central region, $|\eta|<1,$ where $\eta$ is the jet rapidity.  We
only include the transverse energy within the azimuthal region
$|\Delta\phi|<1,$ so in a two-jet event the $E_T$ of the other jet
should not contribute to the pedestal of this jet.  We see that for the
direct events, in which both light-cone momentum fractions are large so
there is no significant underlying event, the multiple scattering makes
very little difference.  On the other hand, for twice-resolved events,
in which both photons have remnants, it makes a lot more difference.
The jet pedestal is sufficiently raised to increase a jet's transverse
momentum by about 200 MeV.  While this may not seem a large shift in
absolute terms, the jet cross-section actually decreases by about 25\%
in going from 3 to 3.2 GeV, giving rise to the correction predicted by
\GGphojet\ in Fig.~\ref{GGMHSpt}b.  Because of the strong correlation
between the shape of the jet and the shift in the cross-section, it
seems hopeful that the models can be constrained by data, to provide a
reliable unfolding of these effects.  Indeed this is already in progress
at HERA.

Of course, one would like to make a more direct measurement of the
nature of the underlying event, to differentiate between soft and
multiple-hard interaction models.  However it is extremely difficult to
find event features that are unambiguous in this respect, as one's
na\"\i ve picture of four jets in back-to-back pairs does not survive
hadronization and realistic jet definition, so one is forced to look in
the hard tails of distributions that are usually not well predicted
anyway.  Nevertheless, if direct evidence of multiple hard interactions
could be found it would be extremely important for our understanding of
photon and hadron collisions as a whole, and it is certainly worth
continuing to search for such signals.

\subsection{Heavy quarks}
\label{GGhq}

In this study we have compared five different generators of charm
production in two-photon events: Vermaseren; \GGpythia; \GGherwig;
\GGgghv01\cite{GGAFaleph,GGAFdkzz}; and \GGminijet\cite{GGAFtopazmc}.
All of these can generate the direct process, but only the last 4 the
resolved process.  For the comparison we chose the same parameters for
each model: charm mass=1.7 GeV; minimum $\mathrm{c\bar{c}}$ invariant
mass=4.0 GeV; beam energy=85.0 GeV; and the GRV parton distribution
set (for the resolved process).

\begin{figure}
\begin{center}
\epsfig{file=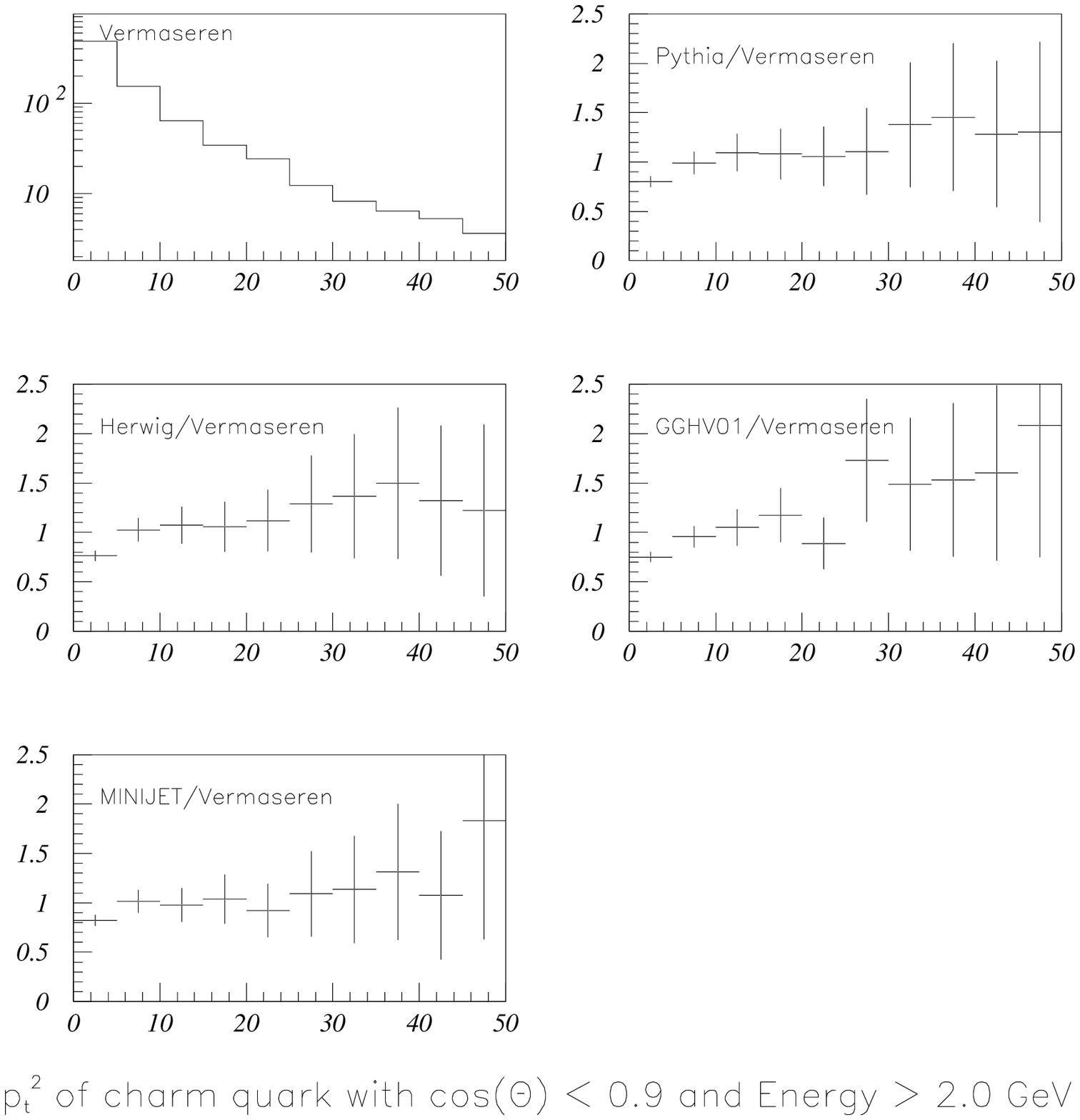,width=1.0\textwidth,height=8.5cm}  
\vspace*{-10mm}
\end{center}
\caption{{\it Comparison of five generators of the direct production
    of charm quarks in two photon collisions. The top left histogram
    is the distribution for the Vermaseren generator. Other generators
    are shown as a ratio to the Vermaseren distribution in remaining
    plots.}}
\label{GGAFfig1}
\end{figure}

\begin{figure}
\begin{center}
\epsfig{file=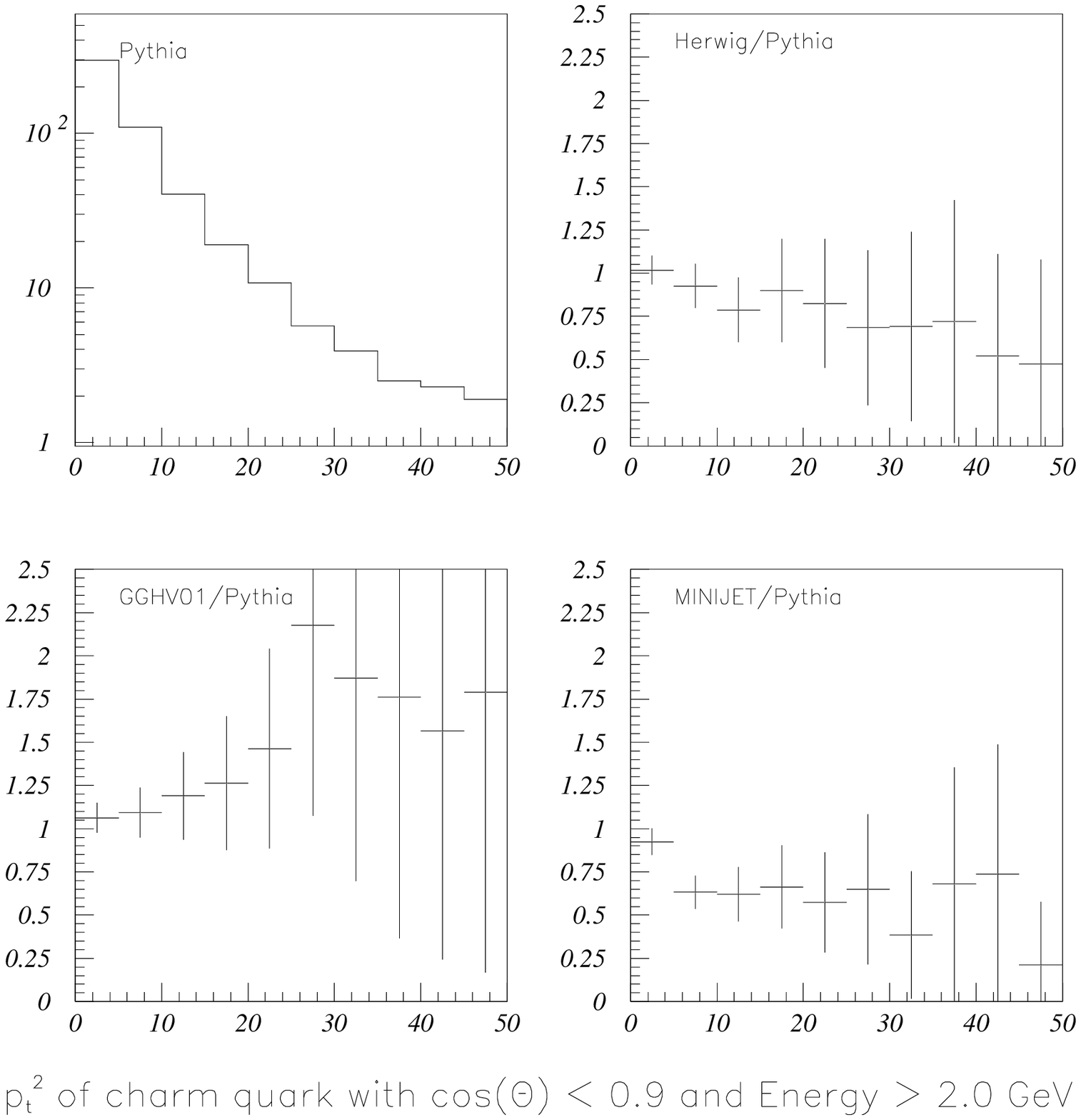,width=1.0\textwidth,height=8.5cm}  
\vspace*{-10mm}
\end{center}
\caption{{\it Comparison of four generators of the production of charm
    quarks via the single resolved process in two photon collisions.
    The top left histogram is the distribution for the \GGpythia\ 
    generator. Other generators are shown as a ratio to the \GGpythia\ 
    distribution in remaining plots.}}
\label{GGAFfig2}
\end{figure}

\begin{figure}
\begin{center}
\epsfig{file=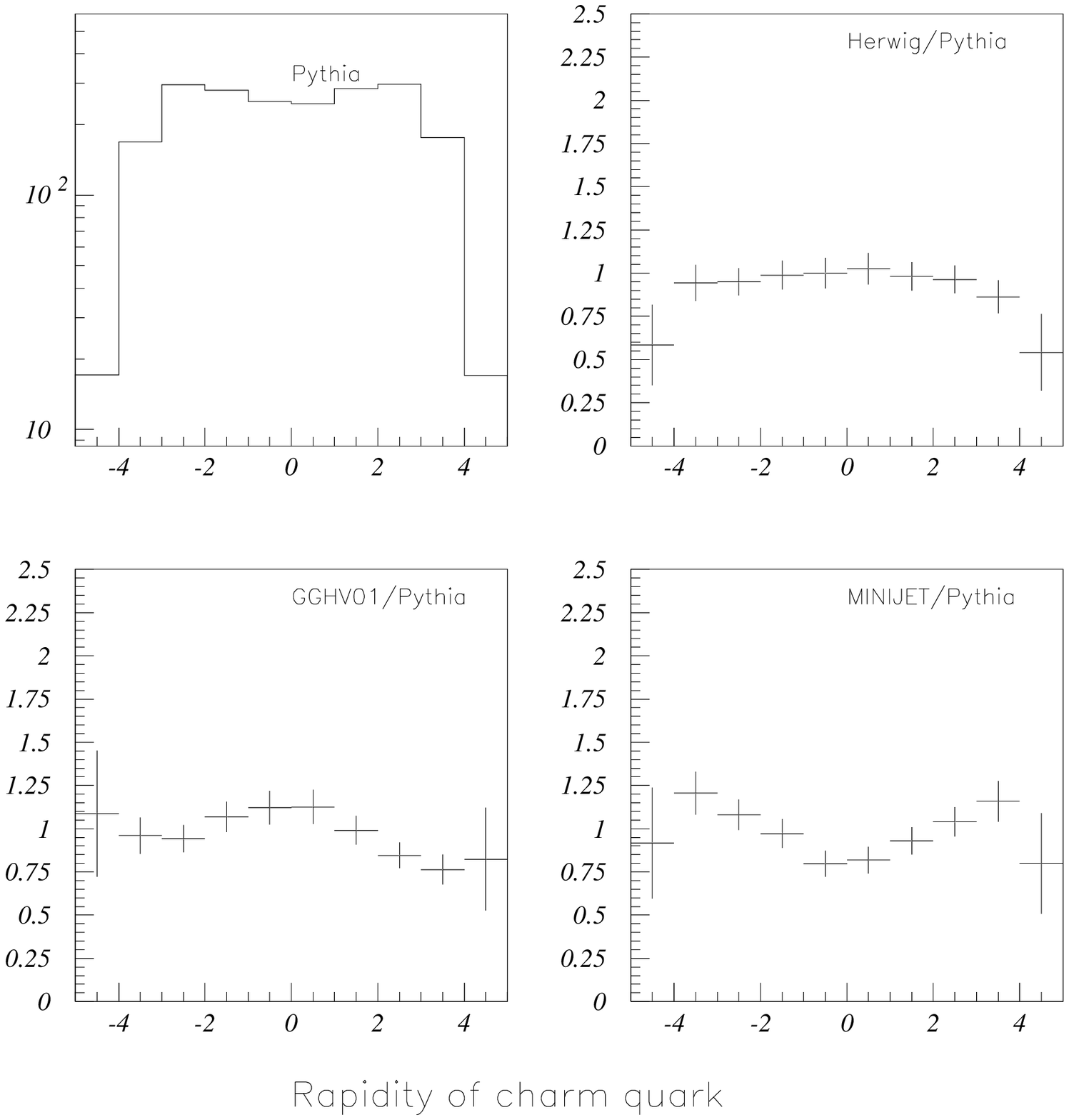,width=1.0\textwidth,height=8.5cm}  
\vspace*{-10mm}
\end{center}
\caption{{\it Comparison of four generators of the production of charm
    quarks via the single resolved process in two photon collisions.
    The top left histogram is the distribution for the \GGpythia\ 
    generator. Other generators are shown as a ratio to the \GGpythia\ 
    distribution in remaining plots.}}
\label{GGAFfig3}
\end{figure}

We generated 10000 events with each program and compared the following
distributions: scattered electron energy; final state invariant mass;
energy, $p_t$ and rapidity of the charm quark; $p_t^2$ of charm quarks
with $\cos(\theta)<0.9$ and energy $>2.0$ GeV.  In most cases the
differences between the generators turn out to be rather small, so
only a small selection are shown here
(Figs.~\ref{GGAFfig1}--\ref{GGAFfig3}).  Since they are so similar, we
emphasize the differences by plotting the absolute distribution for
only one generator (Vermaseren/\GGpythia\ for direct/resolved
production) and showing the results for the other generators as a
ratio to it.

It can be seen that in the case of direct production the programs
produce very consistent results. It should be noted that the
Vermaseren generator is a full matrix element calculation while all
the other generators involve a convolution of some version of the
Equivalent Photon Approximation luminosity function with a
cross-section for real photons.  Thus it should be considered the
correct answer to which the other calculations should be compared.

For the resolved process there is a larger variation between the
generators. Unfortunately at this stage it is not clear which if any
of these should be regarded as standard. The distribution that most
clearly shows the differences is that of the rapidity of the charm
quark (Fig.~\ref{GGAFfig3}), where the \GGminijet\ generator produces
quarks that are more forward peaked, while \GGgghv01\ produces more
quarks in the central region. \GGpythia\ and \GGherwig\ are fairly
similar and lie somewhere in between.

Experimental charm measurement is generally restricted to the central
region, with rapidity less than about 2.  In Table~\ref{GGAFtab1}, we
show the selection efficiency of an imaginary experiment with 100\%
tagging efficiency for all charm quarks within the acceptance cuts,
$\cos(\theta)<0.9$ and $p_\perp>2.0$ GeV.  The differences in the
rapidity distribution for resolved events are directly reflected in
the selection efficiencies.  One can see that an experiment using
\GGminijet\ to correct their data would `measure' a resolved
cross-section almost a factor of two larger than one using \GGgghv01,
with \GGpythia\ and \GGherwig\ lying between the two.  It is therefore
clearly a high priority to understand where these differences arise
and in particular whether the spread represents a genuine uncertainty
in the measurement or is simply an indication that we should not trust
one or some of the models.  Unfortunately little progress has been
made with this during the workshop.
\begin{table}
 \begin{center}
  \begin{tabular}{|c|c|c|}
  \hline
  Generator                   & Direct & Resolved \\
  \hline
  Vermaseren                  &  42 & -  \\
  \GGgghv01\                  &  39 & 28  \\
  \GGpythia\                  &  39 & 24  \\
  \GGherwig\                  &  40 & 21  \\
  \GGminijet\                 &  37 & 16  \\
  \hline
  \end{tabular}
  \caption{{\it Percentage of events passing `acceptance' cuts.}}
  \label{GGAFtab1}
  \end{center}
\end{table}

Comparisons with data for the event features, rather than just the total
cross-section, should also help to resolve this discrepancy.  Within the
limited statistics (33 events) Aleph found that \GGgghv01\ gave a better
description of data than the Vermaseren generator alone\cite{GGAFaleph}.

\section{Description of programs}
\label{GGmainprog}

In this section we detail some of the event generators for
$\gamma\gamma$ physics, concentrating mainly on those in which there
has been significant development during this workshop. Other programs
are commented on and described briefly in section~\ref{GGothers}.

\subsection{Ariadne}

\begin{tabbing}
{\bf Program location:} \= \kill
{\bf Version:}          \> 4.07 of 15 August 1995 \cite{GGLLar}    \\
{\bf Author:}           \> Leif L{\"o}nnblad                       \\
                        \> NORDITA, Blegdamsvej 17,                \\
                        \> DK 2100 Copenhagen {\O}, Denmark        \\
                        \> Phone: + 45 -- 35325285                 \\
                        \> E-mail: leif@nordita.dk                 \\
{\bf Program size:}     \> 12839 lines                             \\
{\bf Program location:} \> http://surya11.cern.ch/users/lonnblad/ariadne/
\end{tabbing}

The \GGariadne\ program implements the Dipole Cascade Model (DCM)
for QCD cascades\cite{GGLLdcm1,GGLLdcm2}. In this model the emission
of a gluon $g_1$ from a $q\bar{q}$ pair created in an $e^+e^-$
annihilation event can be described as radiation from the colour
dipole between the $q$ and $\bar{q}$. A subsequent emission of a
softer gluon $g_2$ can be described as radiation from two independent
colour dipoles, one between the $q$ and $g_1$ and one between $g_1$
and $\bar{q}$. Further gluon emissions are given by three independent
dipoles etc. In this way, the end result is a chain of dipoles, where
one dipole connects two partons, and a gluon connects two dipoles.
This is in close correspondence with the Lund string picture, where
gluons act as kinks on a string-like field stretched between the
$q\bar{q}$ pair.

Further details of how \GGariadne\ generates the QCD cascade in
$e^+e^-$-annihilation can be found in Ref.~\cite{GGLLar} and
elsewhere in these proceedings\cite{GGLLqcdgen}. Here only the parts
relevant for e$\gamma$ DIS and high-$p_\perp$
$\gamma\gamma$ is presented.

The treatment of DIS is very similar to $e^+e^-$\cite{GGLLdcm3}
gluon emission is described in terms of radiation from the colour
dipole stretched between the quark, struck by the electroweak probe,
and the photon remnant. The difference is that, while $q$ and
$\bar{q}$ are both point-like in the case of $e^+e^-,$ the photon
remnant in DIS is an extended object. This results in an extra
suppression of radiation in the remnant direction.

The suppression depends on the transverse size of the remnant, which is
taken to be inversely proportional to its intrinsic transverse
momentum $k_{i\perp}$. In anomalous events where $k_{i\perp}$ is
larger, the suppression is therefore smaller than in
VMD--like events. Also in events where the target photon is
significantly off-shell, the transverse size is taken to be the
inverse of the maximum of $k_{i\perp}$ and the photon
virtuality. Similarly in the case of events with $Q^2\ll W^2,$ where
gluons may be radiated with transverse momentum larger than $Q,$ also
the struck quark may be treated as extended with a transverse size
$\propto 1/Q$.

The DCM is evidently very different from conventional initial-state
parton shower models. Tracing emissions ``backwards'' from the struck
quark as in an initial-state shower, these would be unordered in
transverse momentum, while in a program like \GGpythia\ these
emissions would be ordered in falling transverse momentum. In the DCM,
the transverse momentum of gluons far away from the struck quark,
i.e.\ close to the remnant, can be much larger than in an
initial-state parton shower.

The first emission in the dipole cascade is corrected to match the
${\cal O}(\alpha_S)$ matrix element for gluon emission. The
gamma-gluon fusion diagram is, however, not yet included for the
$e\gamma$ DIS, because of technical difficulties in the current
interface to \GGpythia.

The interface to \GGpythia, described in the report from the QCD
generator group\cite{GGLLqcdgen}, also enables \GGariadne\ to
generate high-$p_\perp$ $\gamma\gamma$ scattering. Here, \GGpythia\
is used to generate the hard sub-process. The outgoing partons and
remnants are then connected with dipoles, which are allowed to
radiate, restricting the transverse momentum of the emissions to be
smaller than that of the hard interaction and treating all remnants as
extended objects as in DIS.

For $\gamma\gamma$ it is also possible to run \GGariadne\ with
\GGpythia\ using multiple interactions. However, this part of the
interface is still preliminary, and more studies are needed.

\subsection{GGHV01}

\begin{tabbing}
{\bf Program location:} \= \kill
{\bf Version:}          \>  Version 1.0 of 1/11/1994                          \\
{\bf Author:}           \>  M.~Kr\"amer, P.~Zerwas (DESY),                    \\
                        \>  J.~Zunft (formerly of DESY, no longer in HEP) and \\
                        \>  A.Finch,                                          \\
                        \>  School of Physics and Materials,                  \\
                        \>  University of Lancaster,                          \\
                        \>  Lancaster LA1 4YB United Kingdom                  \\
                        \>  Phone: (+44) 1524 593618                          \\
                        \>  E-mail: A.Finch@lancaster.ac.uk                   \\
{\bf Program size:}     \>  Interface routines:1148 lines                     \\
                        \>  Physics routines:  2639 lines                     \\
                        \>  Integration package (BASES): 4432 lines           \\
{\bf Program location:} \>  ftp://lavhep.lancs.ac.uk/
\end{tabbing}

\GGgghv01\ was developed by the above authors as a Monte Carlo
implementation of the calculation in Ref.~\cite{GGAFdkzz} of heavy
flavour quark production in gamma gamma collisions at next to leading
order. It can either produce direct or single resolved events at a
range of beam energies, but is restricted to real photons only.

The program is not a complete NLO generator. It does generate the
$2\rightarrow3$ process according to the NLO matrix elements, which is
cut off against soft and collinear divergences. The remaining events
are, however, approximated with $2\rightarrow2$ processes generated
only to leading order, but rescaled so that the total NLO
cross-section is reproduced.

During the initialization stage the routine {\tt FANDK} calculates the
factor by which the $2\rightarrow2$ process should be increased to
achieve the correct cross-section. In order to do this it needs to
know beforehand the total cross-section. This it finds from a
parameterization of the total cross-section from
Ref.~\cite{GGAFdkzz}. This is a quick method but less flexible than
recalculating the total cross-section from scratch. The latter
approach may be adopted in a later version.  The chief criticism of
this approach is that there is just one global correction factor
whereas a more sophisticated approach would be to calculate a
different factor for different regions of phase space.

The main event generation loop uses the coupled routines BASES and
SPRING\cite{GGAFBASES} which together constitute a general purpose
integration and event generation package. The input to both routines
is a function to be integrated as a function of n random variables. In
this case the routine {\tt GENHVY}. This routine uses the input random
numbers to first pick whether to generate a 2 or 3 body event, it then
calculates the fraction of the energy of the incoming beam electrons
taken by each photon. In the resolved photon case the fraction of the
photon's energy taken by a gluon is also found using the Gl\"uck Reya
and Vogt next to leading order parton distribution
functions\cite{GGRE-GRV92b}. Momenta for the outgoing particles are
then chosen using the subroutine {\tt RAMBO}. The weight for the event
is found using routines based on the NLO calculation of
Ref.~\cite{GGAFdkzz} multiplied by the phase space weight from
RAMBO\@. The 2 body (`Born term') contributions is additionally scaled
by the factor calculated in the initialization step as described
above.

During the BASES stage the program first optimizes a `map' of the
function to provide efficient calculation, and then calculates the
total cross-section (for \GGgghv01\ this simply provides a check as
the total cross-section is already known). During the SPRING stage the
function is repeatedly called to generate unit weight events which are
put in the {\tt LUJETS} common for fragmentation by the {\tt LUEXEC}
routine from the \GGjetset\ package. The map generated by BASES is
written out to a file, which means the program can later be run from
the SPRING step only. In this case it reads back the map produced in a
previous run and by simply varying the random number seed a fast event
generation can be achieved..

\subsection{GGPS1, GGPS2}
\label{GGGGPS12}
\begin{tabbing}
{\bf Program location:} \= \kill
{\bf Version:}          \>  1.0 \cite{GGKEKjets}             \\
{\bf Author:}           \> T. Munehisa, K. Kato, D. Perret-Gallix \\
                        \> Phone:
                          +81-552208584 (TM),
                          +81-333421264 (KK), \\
                        \> +33-50091600, +41 22 7676293 (DPG)   \\
                        \> E-mail: munehisa@hep.esb.yamanashi.ac.jp,   \\
                        \> kato@sin.kogakuin.ac.jp, perretg@cernvm.cern.ch \\
{\bf Program size:}     \> 2213 (\GGggpsone) and 2841 (\GGggpstwo) lines   \\
{\bf Program location:} \> ftp://lapphp8.in2p3.fr/pub/keklapp/ggps/ggps.tar.gz
\end{tabbing}

This generator simulates jet productions in two-photon process based
on the leading-log (LL) parton shower (PS) technique\cite{GGKEKll}.
Two cases are separately treated, namely, the deep inelastic
scattering of the photon (\GGggpsone) and the scattering of two
quasi-real photons (\GGggpstwo)\cite{GGKEKjets}. Both processes begin
by the PS space-like evolution, then the hard scattering of partons
takes place, followed by the time-like evolution of the final state
partons.

The non-singlet quark distribution in the photon, $q_{NS}(x,Q^2),$
obeys\cite{GGKEKth}
\begin{equation}
  {d q_{NS}(x,Q^2) \over d \ln Q^2} = {\alpha_s\over 2\pi}
\int_x^1 {dy \over y} P_{qq}^{(0)}(x/y) q_{NS}(y,Q^2)
 +{\alpha\over 2\pi} k_{NS}^{(0)}(x), \label{GGDPGeq:AP}
\end{equation}
The inhomogeneous term, $ k_{NS}^{(0)}(x),$ is proportional to
$x^2+(1-x)^2$.
$\alpha$ is the QED coupling constant, the QCD coupling constant
$\alpha_s$
is defined as: $ \alpha_s(Q^2)=4\pi/(\beta_0 \ln Q^2/\Lambda^2)
 =\alpha_0/(\ln Q^2/\Lambda^2)$ with $\beta_0 = 11-(2/3)N_F$.
$N_F$ is the number of flavours. Eq.~\ref{GGDPGeq:AP} can be brought
into the integral equation
\begin{equation}
q_{NS}(x,Q^2) = \int_x^1 {d y \over y} K_{NS}^{(0)}
(x/y,\overline{s})q_{NS}(y,Q^2_0) \nonumber \\
  + \int^{Q^2}_{Q_0^2}
  {d K^2 \over K^2} \int_x^1 {d y \over y} K_{NS}^{(0)}(x/y,\eta(K^2))
{\alpha \over 2\pi } k_{NS}^{(0)}(y),  \label{GGDPGeq:res}
\end{equation}
where $ \overline{s}\equiv\ln (\alpha_s(Q^2_0)/\alpha_s(Q^2))
= \ln (\ln(Q^2/\Lambda^2)/\ln (Q^2_0/\Lambda^2))$
and $\eta =\ln (\ln(Q^2/\Lambda^2)/\ln(K^2/\Lambda^2))$.
The first term represents the vector meson dominant part (VMD) while
the second corresponds to the perturbative photon.
Here $K_{NS}^{(0)}(x,\overline{s})$ is the QCD kernel
function\cite{GGKEKksy} defined by the inverse Mellin transformation
\begin{equation}
 K_{NS}^{(0)}(x,\overline{s})=\int_{r_0-i \infty}^{r_0+i \infty}
{dn \over 2\pi i} x^{-n} e^{\alpha_0 d(n)\overline{s}},
\end{equation}
where $d(n)$ is the moment of $P_{qq}^{(0)}(x)/(2\pi)$. \par
The singlet distribution is handled in the similar manner.

Due to the inhomogeneous term in Eq.~\ref{GGDPGeq:AP}, the total energy
is conserved {\em only if\/} the photon energy is included,
\begin{equation}
 \int_0^1 dx x \lbrack \sum_f
( q_f(x,Q^2)+\overline{q}_f(x,Q^2))+ G(x,Q^2) \rbrack
  = \int^{Q^2}_{Q_0^2} {d K^2 \over K^2}
{\alpha \over 2\pi }\int_0^1 d x x  k_{NS}^{(0)}(x).
\label{GGDPGeq:eng}
\end{equation}
In the generation the right-hand side is used as
{\em the weight of the event}.

Let us summarize the main steps present in the algorithm
generating the partons in the course of the evolution.
\begin{enumerate}
\item Selection of a $Q^2$.
\item Calculation of the energy of the VMD part (independent of
$Q^2$) and of
the energy of the perturbative photon part by using
Eq.~\ref{GGDPGeq:eng}.
The sum of these energies is used as the weight of the event.
\item Selection of the actual process, either VMD or perturbative
photon,
    according to the ratio of energies.
\item If VMD is chosen, the usual QCD evolution
    from $Q^2_0$ (the cut-off momentum) to $Q^2$ is performed.
\item In the case of a perturbative photon, the virtual mass
squared $K^2$ according to the probability $d K^2/K^2$ is determined.
Then the flavour of the partons are selected according to the ratio
of charges
squared.
\item For each quark or anti-quark, the usual
QCD evolution from $K^2$ up to $Q^2$ is performed.
\end{enumerate}
This algorithm is common to the deep inelastic photon scattering
and to the
two quasi-real photon scattering with large $p_T^2$.
The hard scattering parts are, however, different.
In quasi-real photon collision the initial states $q\hbox{-}q,$
$q\hbox{-}\overline{q},$ $q\hbox{-}G,$ $G\hbox{-}G,$ $q\hbox{-}\gamma$
$G\hbox{-}\gamma$ and $\gamma\hbox{-}\gamma$ are taken into account
for the hard scattering, although in the virtual-quasi-real case only
the $\gamma^*$-$q$ and $\gamma^*$-$G$ subprocesses are considered.
\par
The reference cross-section used to select events is the differential
cross-section:
\begin{equation}
d\sigma^0/dp_T^2 = \pi/p_T^4.
\label{GGDPGeq:1}
\end{equation}
The ratio of the hard cross-section to
Eq.~\ref{GGDPGeq:1} is counted as the weight
of the event.
The argument of $\alpha_s$ in the hard cross-section is set to be
$ p_T^2 ,$ because there are some processes whose cross-section is
dominated by the $t$-channel contribution.
After the time-like evolution of the produced
partons has been performed, the energy of the initial photons is fixed
and thus the four-momenta of all partons are determined.

Events are generated with a weight whose maximum is unknown at the
beginning of a run. A maximum value is arbitrary fixed by the user,
overweight
events are rejected and counted. At the end of the run, if the number
of
overweight events is too large, compromising the generation accuracy,
the user
must increase the maximum weight, reducing consequently the generation
efficiency.

Finally the generated partons are hadronized
following the
\GGjetset\cite{GGTSpyje} mechanism. Colour flows are properly matched
and adopt the \GGjetset\ definition. Two types of colour singlet exist:
a string beginning and ending with quarks possibly embedding one
or many gluons or a closed colour loop of gluons.

\subsection{HERWIG}
\label{GGherwig}
\newcommand{\MHSgtap}{\raisebox{-.47ex}{\rlap{$\sim$}} \raisebox{.47ex}{$>$}}
\newcommand{\MHSltap}{\raisebox{-.47ex}{\rlap{$\sim$}} \raisebox{.47ex}{$<$}}
\begin{tabbing}
{\bf Program location:} \= \kill
{\bf Version:}          \> 5.8d of October  1995 \cite{GGMHSherwig}
                      (and 5.9  of December 1995) \\
{\bf Authors:}          \> G. Marchesini$^1,$ B.R. Webber$^2,$
                           G. Abbiendi$^3,$ I.G. Knowles$^4,$      \\
                        \> M.H. Seymour$^5,$ L. Stanco$^3$         \\
                        \> $^1$Dipartimento di Fisica, Universita di Milano. \\
                        \> $^2$Cavendish Laboratory, University of Cambridge. \\
                        \> $^3$Dipartmento di Fisica, Universita di Padova. \\
                        \> $^4$Department of Physics and Astronomy, University
                           of Glasgow.                             \\
                        \> $^5$Theory Division, CERN.               \\
                        \> E-mail: webber@hep.phy.cam.ac.uk,
                                   knowles@v6.ph.gla.ac.uk,        \\
                        \>         seymour@surya11.cern.ch.        \\
{\bf Program size:}     \> 15500 lines                             \\
{\bf Program location:} \> http://surya11.cern.ch/users/seymour/herwig/
\end{tabbing}

\GGherwig\ is a general-purpose QCD Monte Carlo event generator for
simulating {\bf H}adron {\bf E}mission {\bf R}eactions, {\bf W}ith
{\bf I}nterfering {\bf G}luons.  Its general design philosophy is to
provide as complete as possible an implementation of perturbative QCD,
combined with as simple as possible a model of non-perturbative QCD\@.
It does this uniformly for a very wide range of processes, allowing
the parameters to be fitted in one reaction, principally
$\mathrm{e^+e^-}$ annihilation, and applied to other reactions.
Although \GGherwig\ has been capable of simulating $\gamma\gamma$
collisions for some time, there were many technical deficiencies,
which practically prevented experiments from using it.  Many of these
have been rectified during the workshop, resulting in the preliminary
versions 5.8a, 5.8c and 5.8d.  However work is still ongoing, and some
of the features described below will not be available until the full
version release towards the end of this year.  Since much of
\GGherwig\ is described in Ref.~\cite{GGLLqcdgen}, we concentrate on
the additional features relevant to $\gamma\gamma$ collisions here.
Original references can be found in Ref.~\cite{GGMHSherwig}.

Event generation proceeds much like the general case described in the
introduction.  The equivalent photon approximation (EPA) is used,
correctly generating the $P^2$ dependence.  Since this means that the
photons no
longer collide head-on, they are boosted to their centre-of-mass frame,
where the remainder of the event generation is performed.  At the end of
the event they are automatically boosted back to the lab frame.

Since the user can control the $P^2$ range generated (through the
variables \verb+Q2WWMN+ and \verb+Q2WWMX+), it is in principle possible
to generate all event classes using the EPA\@.  However, to provide a more
accurate description of high-$Q^2$ tagged events, these are described as
deep inelastic scattering, $\mathrm{e\gamma\to e+}$hadrons including the
full electron kinematics, rather than as $\gamma^*\gamma\to$hadrons
using the EPA.

The treatment of deep inelastic lepton-photon scattering is essentially
identical to that of lepton-hadron, except for the inclusion of the
point-like photon-quark coupling in the initial-state parton shower.
The hard process is generated as
$\mathrm{eq\to eq}$ according to whichever parton distribution function
is selected from \GGpdflib.  This is controlled by the variables
\verb+AUTPDF+ and \verb+MODPDF+, which hold the `author group' and set
number respectively, for example \verb+GRVph+ and \verb+3+ for the
leading-order photon set of Gl\"uck, Reya and Vogt.

The outgoing quark produces a parton shower exactly like that in the
final state of $\mathrm{e^+e^-}$ annihilation, described in
Ref.~\cite{GGLLqcdgen}.  The incoming quark also produces a parton
shower, which is generated using the `backward evolution' algorithm.
One can imagine this as an evolution in the factorization scale from
the large scale of the hard process down towards zero~--- as it is
reduced, more and more radiation is resolved, i.e.~transferred from
the evolution of the parton distribution function to the coefficient
function.  The inhomogeneous term in the evolution equation
corresponds directly to the addition of a $\gamma\to\mathrm{q\bar{q}}$
vertex, which is straightforwardly included by considering the photon
to be an additional parton type with a delta-function distribution.
This results in a dynamic separation of events into point-like and
hadronic.  At small and medium $x$ values ($x\MHSltap0.3$), this
separation is similar to that made in the input distribution
functions, demonstrating the self-consistency of the backward
evolution algorithm, but at large $x$ it becomes increasingly
different from them, classifying almost all large-$x$ events as
hadronic, whereas most distribution function sets classify them as
point-like.  The difference can be traced to the fact that \GGherwig\ 
uses a cutoff in transverse momentum to separate the perturbative and
non-perturbative regions, whereas the distribution functions make a
cut on the virtuality of the internal line.  At large $x,$ the two
differ by a factor of $(1-x)$ giving large differences in the
separation scales.  While it could be argued that transverse momentum
is a more physical scale to use, as in the FKP model, a more
conservative approach is to simply say that this indicates a region of
uncertainty, and any analysis that relies on the classification of
this large $x$ region into the two components should be considered
highly model-dependent.

All partons produced by the initial-state cascade undergoing further
parton showering just like any other outgoing parton.

Colour coherence is as important in initial-state cascades as it is in
the final state.  However it is less well understood and only an
approximate treatment is implemented in \GGherwig.  At large $x,$ all
emitted gluons must be soft, the emission pattern is identical to a
final state dipole, and the emission is angular-ordered.  At small
$x,$ rather than restricting soft gluon momenta, we actually `look
inside' the soft gluons, resulting in a different coherence structure.
At `fairly small'~$x,$ this can be approximated by imposing transverse
momentum ordering, which is how it is implemented in \GGherwig, but at
very small $x$ the correct approach is to use angular ordering and
modified emission probabilities.  Although this has never been
implemented as a full Monte Carlo event generator, a partial
implementation was discussed in Ref.~\cite{GGMHSsmallx} and shown to
give similar results to the \GGherwig\ algorithm at the $x$ values
expected at LEP~2 or even HERA\@.  The actual evolution variable used
by \GGherwig\ smoothly interpolates the two regions.

Just as in $\mathrm{e^+e^-}$ annihilation, \GGherwig\ is not capable
of covering the whole of phase-space with its parton shower emission.
This is corrected using the methods proposed in Ref.~\cite{GGMHShard},
which ensure that the {\em hardest\/} emission (which is not
necessarily the first owing to the ordering of opening angles) agrees
with the exact matrix element (the sum of higher-order resolved,
$\mathrm{g\to q\bar{q},\;q\to qg},$ and point-like,
$\gamma\to\mathrm{q\bar{q}}$).  It is worth noting that azimuthal
correlations\cite{GGLLggphys} are correctly included within the
`dead-zone' region of $x_p$ and $z_p$ defined in
Ref.~\cite{GGMHSglasgow}, but not within the parton shower itself.

After the parton cascade, the system is hadronized according to the
cluster model.  For outgoing partons, this is identical to
$\mathrm{e^+e^-}$ annihilation, described in Ref.~\cite{GGLLqcdgen},
but for hadronic events, the photon remnant is treated specially.  It
is given a limited transverse momentum of width \verb+PTRMS+, and the
cluster that contains it is given a Gaussian mass-squared
distribution, resulting in a limited $p_t$ spectrum of produced
hadrons, as expected for such a soft object.  An `underlying event'
model is provided, which increases the energy released during the
break-up of the remnant, but it seriously overestimates HERA data on
DIS at small $x,$ and it is recommended that it be switched off by
running process \verb+IPROC=19000+ rather than
\verb+9000+\footnote{This applies to versions 5.8d onwards.  In
  earlier versions, the two processes would be expected to bracket the
  correct result.}.

Untagged and low-$Q^2$ tagged events are generated using the equivalent
photon approximation to split both beams to photons.  One hard process
type is selected from those listed in Ref.~\cite{GGLLqcdgen}.  At present
there is no facility to mix events of different types in the same run,
although it is clear that this would improve the utility of the program
and it is a planned improvement.  Events are generated according to
the leading order matrix elements for $2\to2$ scattering, and
initial-state and final-state showers are added exactly as in deep
inelastic scattering.  Another current problem is that the phase-space
cuts accessible to the user are completely different for direct and
resolved processes, making it difficult to generate both uniformly.
Fixing this is another planned improvement.

In all event classes except deep inelastic scattering, it is recommended
that the soft underlying event be selected.  This models the collision
between the photon remnants as a soft hadron-hadron scattering, which
produces a uniform rapidity plateau of extra hadrons on top of those
from the perturbative event.  This is based on the UA5 minimum bias
model, and is essentially just a parametrization of data.  The soft VMD
scattering process, \verb+IPROC=8000+, uses the Donnachie-Landshoff
cross-section and generates events as in the soft underlying event.

As discussed earlier, one would expect multiple hard scattering to be
an important effect in untagged $\gamma\gamma$ collisions at LEP~2.
Its contribution to the underlying event is also a major source of
uncertainty in the measurement of jets at HERA and it is important to
have several different models of it, to get some estimate of the
uncertainty in the predictions.  Although such scattering is not
included in \GGherwig\ at present, there is an interface to the
\GGjimmy\ Generator\cite{GGMHSjim} (which can also be obtained from
the web-page listed above).  However, there is not yet a smooth
transition between hard and soft multiple scattering, and \GGjimmy\ 
and s.u.e.\ should be considered mutually exclusive at present.

The default parameter set that comes with \GGherwig\ is tuned to
OPAL data in the case of parameters that affect
$\mathrm{e^+e^-}$ annihilation, but are simply theoretically prejudiced
guesses for the others.  At present the HERA experiments are involved
with tuning the additional parameters and these will be included in
future releases, hopefully improving the predictivity for $\gamma\gamma$
physics.

\subsection{PHOJET}
\label{GGphojet}

\begin{tabbing}
{\bf Program location:} \= \kill
{\bf Version:}          \> 1.04 of 20 October 1995 \cite{GGRE-Engel95a}\\
{\bf Author:}           \> Ralph Engel                             \\
                        \> Institute of Theoretical Physics        \\
                        \> University Leipzig                      \\
                        \> Augustusplatz 10, D-04109 Leipzig, Germany\\
                        \> Phone: + 49 -- 341 - 97 32444            \\
                        \> E-mail: eng@tph200.physik.uni-leipzig.de\\
{\bf Program size:}     \> 31000 lines                     \\
{\bf Program location:} \> http://www.physik.uni-leipzig.de/\verb+~+engel/phojet.html
\end{tabbing}

\GGphojet\ is a minimum bias event generator for hadronic
pp, $\gamma$p and $\gamma\gamma$ interactions. 
The interactions are described within the Dual Parton Model 
(DPM)\cite{GGRE-Capella94a} in terms of reggeon and pomeron exchanges. 
The realization of the DPM with a hard and a soft
component in \GGphojet\ is similar to the event generator {\sc
  Dtujet}-93\cite{GGRE-Aurenche92a,GGRE-Aurenche94a}.
Regge arguments are combined with perturbative QCD to get an almost complete
description of the leading event characteristics.
Special emphasis is taken on diffractive and soft interactions. Soft and hard
interactions are unitarized together leading to the possibility to have 
multiple soft and hard scatterings in one event.

In the following, some comments on LEP~2 specific aspects are given.
In the model\cite{GGRE-Engel95a}, 
the dual nature of the photon is taken into account
by considering the physical photon state as a
superposition of a "bare photon" and virtual hadronic states having the
same quantum numbers as the photon.
Two generic hadronic states $| q\bar q\rangle$ and
$| {q\bar q}^\star\rangle$ have been introduced to describe the hadronic
piece of the photon. The low-mass state $| q\bar q\rangle$ corresponds
to the superposition of the vector mesons $\rho,$ $\omega$ and $\phi$
and a $\pi^+\pi^-$ background. The state $| {q\bar q}^\star\rangle$ is
used as an approximation for hadronic states with higher masses.
The physical photon reads
\begin{equation}
|\gamma \rangle = \sqrt{Z_3}\ |\gamma_{\mbox{\scriptsize bare}} \rangle +
\frac{e}{f_{q\bar{q}}} \ |q\bar{q} \rangle +
\frac{e}{f_{q\bar{q}^\star}} \ |q\bar{q}^\star \rangle{}.
\end{equation}
The interaction of the hadronic states via pomeron/reggeon exchange is
subdivided into processes involving only {\em soft\/}
processes and all the other processes with at least one large momentum
transfer ({\em hard\/} processes) by applying
a transverse momentum cutoff $p_\perp^{\mbox{\scriptsize cutoff}}$ to
the partons.
On Born-graph level, for example, the photon-photon cross-sections is built
up by: {\bf (i)} soft reggeon and pomeron exchange,
{\bf (ii)} hard resolved photon-photon interaction,
{\bf (iii)} single direct interactions, and
{\bf (iv)} double direct interactions.
The soft pomeron cross-sections is parametrized using Regge theory.
The hard cross-sections are calculated within the QCD Parton Model using
lowest order matrix elements. For soft processes, photon-hadron 
duality is assumed.  The energy-dependence of the reggeon and
pomeron amplitudes is assumed to be the same for all hadronic processes.
Therefore, data on hadron-hadron and photon-hadron cross-sections
can be used to determine the parameters necessary to describe soft 
photon-photon interactions.

The amplitudes corresponding to the one-pomeron exchange between the
hadronic fluctuations are
unitarized applying a two-channel eikonal formalism similar to
Ref.~\cite{GGRE-Aurenche92a}{}. 
The probabilities $e^2/f^2_{q\bar q}$ and $e^2/f^2_{q\bar{q}^\star}$ 
to find a photon in one of the generic hadronic 
states, the coupling constants to the reggeon and pomeron, and the 
effective reggeon and pomeron intercepts cannot be 
determined by basic
principles. These quantities are treated as free parameters and determined
by cross-section fits\cite{GGRE-Engel95a}{}. Once the parameters are fitted,
the model allows for predictions on photon-photon collisions without 
new parameters.

The probabilities for the different partonic final
state configurations are calculated from the discontinuity of the
scattering amplitude (optical theorem). Using the 
Abramovski-Gribov-Kancheli cutting
rules\cite{GGRE-Abramovski73} the cross-section for
graphs with $k_c$ soft pomeron cuts, $l_c$
hard pomeron cuts, $m_c$ triple- or loop-pomeron cuts, and $n_c$
double-pomeron are estimated.
For pomeron
cuts involving a hard scattering, the complete parton kinematics and
flavours/colours are sampled according to the Parton Model using a method
similar to Ref.~\cite{GGRE-Hahn90}, extended to direct processes. 
For pomeron cuts involving parton configurations without a large
momentum transfer, the partonic interpretation of the Dual Parton Model
is used: photons or mesons are split into a quark-antiquark pair whereas 
baryons are approximated by a quark-diquark pair.
The longitudinal momentum fractions of the soft partons
are given by Regge asymptotics\cite{GGRE-Capella80b}{}.
One obtains for the valence quark ($x$) and diquark ($1-x$)
distribution inside the proton
$ \rho(x) \sim (1-x)^{1.5}/\sqrt{x}$
and for the quark antiquark distribution inside the photon 
$ \rho(x) \sim 1/\sqrt{x (1-x)}$.
\label{x-meson}
For multiple interaction events, the sea quark momenta are sampled from a
$ \rho(x) \sim 1/x $
distribution.
The transverse momenta of the soft partons are sampled from an 
exponential distribution in order to get a smooth transition between
the transverse momentum distributions of the
soft constituents and the hard scattered partons.

In diffraction dissociation or double-pomeron scattering, 
the parton configurations are generated using the
same ideas described above applied to pomeron-
photon/pomeron scattering processes. According to the kinematics 
of the triple- or loop-pomeron graphs, the
mass of the diffractively dissociating systems is sampled from a
 $1/M_D^{2\alpha_\GGREPom(0)}$ distribution. The momentum transfer in
diffraction
is obtained from an exponential distribution with mass-dependent slope
(see Ref.~\cite{GGRE-Engel95a}).
For the parton distributions of the pomeron, the
CKMT parametrization with a hard gluonic 
component\cite{GGRE-Capella95a}
is used. The low-mass part of diffraction dissociation is approximated by 
the superposition of high-mass vector mesons.
In order to take into account the
transverse polarization of quasi-elastically produced
vector mesons, diffractively scattered $\rho,$ $\omega$ and $\phi$ 
are decayed anisotropically.

Finally, the fragmentation of the sampled partonic final states is done
by forming colour neutral strings between the partons according to the 
colour flow. In the limit of many colons in QCD, 
this leads to the two-chain configuration
characterizing a cut pomeron and a one-chain system for a cut reggeon.
In hard interactions the colour flow is taken from the matrix elements
directly\cite{GGRE-Bengtsson84}.
The leading contributions of the matrix elements give a two-chain structure
which corresponds to a cut pomeron.
The chains are fragmented using the Lund fragmentation code \GGjetset\
7.3\cite{GGTSpyje}.

In order to get the LEP~2 kinematics, the complete lepton-photon vertex
for transversally polarized photons is simulated in the program. The 
lepton (anti-) tagging conditions can be specified by the user. However, in
the model only photons with very low virtualities are considered at the moment.
The extension to virtual (and longitudinally polarized)
photons is in progress.

An example input file for a $\gamma\gamma$ run can be in found in the
\GGphojet\ manual\cite{GGRE-Engel95e}.

\subsection{PYTHIA/JETSET}
\label{GGpythia}

\begin{tabbing}
{\bf Program location:} \= \kill
{\bf Versions:}         \> \GGpythia\ 5.720 of 29 November 1995 \cite{GGTSpyje}\\
                        \> \GGjetset\ 7.408 of 23 August 1995    \\
{\bf Author:}           \> Torbj\"orn Sj\"ostrand                  \\
                        \> Department of Theoretical Physics       \\
                        \> University of Lund                      \\   
                        \> S\"olvegatan 14A, S-223 62 Lund, Sweden \\
                        \> Phone: + 46 -- 46 - 222 48 16           \\
                        \> E-mail: torbjorn@thep.lu.se             \\
{\bf Program size:}     \> 19936 + 11541 lines                     \\
{\bf Program location:} \> http://thep.lu.se/tf2/staff/torbjorn/
\end{tabbing}

\GGpythia/\GGjetset\ is a general-purpose generator of high-energy
particle physics reactions. An introduction and a survey of processes
of interest for LEP~2 physics are found in the QCD generators section,
while the full description is in Ref.~\cite{GGTSpyje}.
Here only aspects specific to $\gamma\gamma$ physics will be summarized.
These have been developed together with Gerhard~Schuler, and are
described extensively elsewhere\cite{GGTSgamp,GGTSgamgam}.

To first approximation, the photon is a point-like particle. However,
quantum mechanically, it may fluctuate into a (charged) 
fermion--antifermion pair. The fluctuations 
$\gamma \leftrightarrow \GGTSq\GGTSqbar$ can interact strongly and therefore 
turn out to be responsible for the major part of the 
$\gamma\gamma$ total cross-section. The spectrum of fluctuations 
may be split into a low-virtuality and a high-virtuality
part. The former part can be approximated by a sum over low-mass 
vector-meson states, customarily restricted to the lowest-lying vector 
multiplet. Phenomenologically, this Vector Meson Dominance (VMD) ansatz 
turns out to be very successful in describing a host of data. The 
high-virtuality part, on the other hand, should be in a perturbatively 
calculable domain. Based on the above separation, \GGpythia\ distinguishes
three main classes of interacting photons: direct, VMD and anomalous. 

For a $\gamma\gamma$ event, there are therefore three times three
event classes. By symmetry, the `off-diagonal' combinations appear
pairwise, so the number of distinct classes is only six.
These are:
\begin{GGTSenumerate}
\item VMD$\times$VMD: both photons turn into hadrons, and the processes
are therefore the same as allowed in hadron--hadron collisions.
\item VMD$\times$direct: a bare photon interacts with the partons of the
VMD photon.
\item VMD$\times$anomalous: the anomalous photon perturbatively 
branches into a $\GGTSq\GGTSqbar$ pair, and one of these (or a daughter parton 
thereof) interacts with a parton from the VMD photon.
\item Direct$\times$direct: the two photons directly give a quark pair,
$\gamma\gamma \to \GGTSq\GGTSqbar$. Also lepton pair production is allowed,
$\gamma\gamma \to \ell^+\ell^-,$ but will not be considered here.
\item Direct$\times$anomalous: the anomalous photon perturbatively 
branches into a $\GGTSq\GGTSqbar$ pair, and one of these (or a daughter parton 
thereof) directly interacts with the other photon. 
\item Anomalous$\times$anomalous: both photons perturbatively branch 
into $\GGTSq\GGTSqbar$ pairs, and subsequently one parton from each photon 
undergoes a hard interaction.
\end{GGTSenumerate}
In a complete framework, there would be no sharp borders between the
six above classes, but rather fairly smooth transition regions that 
interpolate between the extreme behaviours. However, at our current
level of understanding, we do not know how to do this, and therefore 
push our ignorance into parameters of the model. From a practical point of 
view, the sharp borders on the parton level are smeared out by parton 
showers and hadronization. Any Monte Carlo event sample intended to catch a 
border region would actually consist of a mixture of the three 
extreme scenarios, and therefore indeed be intermediate. 

The main parton-level processes that occur in the above classes are:
\begin{GGTSitemize}
\item The `direct' processes $\gamma\gamma \to \GGTSq\GGTSqbar$ only occur 
in class 4.
\item The `1-resolved' processes $\gamma\GGTSq \to \GGTSq\GGTSg$ and 
$\gamma\GGTSg \to \GGTSq\GGTSqbar$ occur in classes 2 and 5.
\item The `2-resolved' processes $\GGTSq\GGTSq' \to \GGTSq\GGTSq'$ 
(where $\GGTSq'$ may also represent an antiquark), 
$\GGTSq\GGTSqbar \to \GGTSq'\GGTSqbar',$ $\GGTSq\GGTSqbar \to \GGTSg\GGTSg,$ 
$\GGTSq\GGTSg \to \GGTSq\GGTSg,$ $\GGTSg\GGTSg \to \GGTSq\GGTSqbar$ and
$\GGTSg\GGTSg \to \GGTSg\GGTSg$ occur in classes 1, 3 and 6.
\item Elastic, diffractive and low-$p_{\perp}$ events occur in class 1. 
\end{GGTSitemize}
The notation direct, 1-resolved and 2-resolved is the conventional
subdivision of $\gamma\gamma$ interactions. The rest is then called `soft-VMD'. 
The subdivision in \GGpythia\ is an attempt to be more precise and internally
consistent than the conventional classes allow. 

The cross-sections for the above components are based on a number of
considerations. The VMD$\times$VMD ones are derived from a Regge theory
ansatz, with a pomeron plus reggeon form for the total cross-section,
plus parametrizations of elastic and diffractive components\cite{GGTScross}.  
The other five cross-sections are obtained by a direct integration of
parton-level cross-sections above lower cut-offs $k_0 \approx 0.6$~GeV for 
$\gamma \leftrightarrow \GGTSq\GGTSqbar$ fluctuations and 
$p_{\perp\GGTSindx{min}}^{\GGTSindx{anom}} = %
0.7 + 0.17 \log^2(1+E_{\GGTSindx{cm}}/20)$~[GeV]
for QCD processes in the anomalous sector. The latter is a purely 
phenomenological fit based on consistency arguments in the $\gamma$p sector.
It does not include possible eikonalization effects, and would therefore
change in a more complete treatment (studies under way). Taken all together,
one obtains a reasonable description of the total $\gamma\gamma$ cross-section.

The program comes with several parton-distribution sets for the
photon.  The default is SaS~1D\cite{GGRE-Schuler95a}. In view of the
above event classification, the SaS sets have the advantage that the
separation into VMD and anomalous parts is explicit.

Currently only real incoming photons are considered.
Eventually \GGpythia\ should be extended to (moderately) virtual photons.
A separate treatment exists of the DIS region, e$\gamma \to \mbox{e}X,$
i.e.\ with one real and one very virtual $\gamma,$ but this option is less
well developed (especially for parton showers). Furthermore, the program
does not yet generate the spectrum of real photons internally, i.e.\
it is easiest to run for a fixed energy of the $\gamma\gamma$ system.
It is possible to use the varying-energy and weighted-events options of 
the program to obtain a reasonable photon spectrum, however.  

Some main switches and parameters of interest for $\gamma\gamma$ physics
are:
\begin{GGTSitemize}
\item {\tt MSEL} selects allowed processes; a change to 2 would include
also elastic and diffractive VMD events.
\item {\tt MSTP(14)} sets the $\gamma\gamma$ event class among the six
possibilities listed above. The most useful option is {\tt MSTP(14) = 10},
where the six classes are automatically mixed in the proper proportions.
Note that some variables, such as {\tt CKIN(3)}, differ between the event
classes and therefore automatically are reset internally.
\item {\tt MSTI(9)} tells which event class the current event belongs to.
\item {\tt MSTP(55)} and {\tt MSTP(56)} gives choice of parton-distribution
set and library for the photon.
\item {\tt PARP(15)} is the $k_0$ scale, i.e.\ minimum $p_{\perp}$ for
$\gamma \leftrightarrow \GGTSq\GGTSqbar$ fluctuations. 
\item {\tt CKIN(3)} sets $p_{\perp\GGTSindx{min}}$ scale of hard interactions;
for {\tt MSTP(14) = 10} only to be set when studying high-$p_{\perp}$ jet 
production.
\item {\tt PARP(81)} (alternatively {\tt PARP(82)}, depending on {\tt MSTP(82)}
value) sets the $p_{\perp\GGTSindx{min}}$ scale of multiple interactions in 
VMD$\times$VMD events.
\end{GGTSitemize}

\section{Other generators}
\label{GGothers}

Most of the general-purpose QCD event generators described above have only
become available for $\gamma\gamma$ physics during the last year or two.
Therefore most existing analyses have used generators written specifically
for $\gamma\gamma$ physics with little or no overlap with other processes.
This means that they are extremely hard to test thoroughly, except on the
very data they are used to unfold.  Although of course the different
programs have different strengths and weaknesses, they are mainly based on
the same general design.  In this section we describe it and make a few
general comments on when and why it is expected to be adequate or
inadequate.

The principle way in which the older programs differ from the QCD
generators is the lack of a parton shower stage.  A hard scattering is
typically generated according to the leading order matrix element, often
taking great care with their accuracy and including many effects that are
not yet included in the QCD programs.  The final state described by these
matrix elements ranges from a simple quark-antiquark pair in the case of
DIS to up to two jets and two remnants in the case of high-$p_t$
scattering.  These simple partonic states are then given directly to the
\GGjetset\ program, which implements the Lund string fragmentation model.

It may at first sight appear that parton showers are a luxurious frill
that can be added to the model to slightly improve event simulation,
for example by including the small fraction of events in which a third
jet accompanies the two hard jets in an event.  However in the modern
view of hadronization, they are an essential precursor to the
confinement of partons into hadrons, which does not take place
globally between the main few hard partons in an event but locally
between nearby soft partons.  It is only by adopting the
latter view that one can have any expectation that the hadronization
process is universal.  Quite apart from these theoretical issues, as a
practical issue it is certainly the case that the parameters of the
string model need to be retuned to fit to data at different energies
and in different reactions when it is not preceded by a parton shower,
but that coupled with a parton shower a reasonable description of all
current data can be achieved with a single parameter set. This alone
is enough to mean that any model that uses string fragmentation
without a parton shower should be considered descriptive rather than
predictive.

In addition to their r\^ole in setting the initial conditions for the
hadronization process, parton showers are crucial for determining
certain event features, for example the hadronic final state of DIS at
small $x,$ where the predictions of a `QPM model', i.e.\ hard matrix
element plus string fragmentation were found to give an extremely poor
description of the HERA data\cite{GGMHSh1}.

Another closely related difference between the two groups of programs
is in the assumptions made about the photon structure function.  The
older programs generally make an FKP-like\cite{GGLLFKP} separation
into the point-like and hadronic parts, a scheme with strong physical
motivation, but then neglect the QCD evolution in the hadronic part.
While one might suppose that this is unimportant if we only use the
event generator as an unfolding tool, this is not in fact the case
because structure functions do not just evolve by themselves, they do
so by emitting gluons, which have an effect on the structure of the
hadronic final state.  QCD not only predicts that changing $Q^2$
changes the structure function, but also the features of the hadronic
final state.  This should therefore be included in any model used to
unfold data.

One possible solution would be to incorporate parton showers into the
existing programs.  However this is far from straightforward, because the
evolution of a shower is strongly dependent on its initial conditions,
namely on how the hard partons were formed.  For example in DIS, simply
setting up a $\mathrm{q\bar{q}}$ pair and calling \GGjetset\ with its final
state parton shower option switched on will result in emission with
transverse momenta up to $W/2,$ producing far too much hadronic activity.
The correct solution, roughly speaking, is that the current jet produces a
shower with upper scale of order $Q,$ the photon remnant produces one of
order its $p_t$ and the internal line produces an initial-state shower.
There are additional complications in the high-$p_t$ case as there are
contributions that contribute to different event classes depending on their
kinematics, so one must impose additional constraints on the parton shower
in order to avoid double-counting.  Eventually, one realizes that to build
in all the relevant conditions, the parton shower has to know so much about
the hard process that one ends up almost having to write ones own parton
shower algorithm from scratch.

Should we therefore conclude that existing programs are wrong and
existing analyses need to be redone?  No, because in the kinematic
regions explored so far they have several advantages.  Firstly the
range in $W$ has been rather limited so that the move into regions in
which parton showers and QCD effects become important has been
limited.  Secondly for reasonably low $W,$ most of the hadronic event
is actually seen in the detector, so the actual reliance on the models
is rather small.  Thirdly, this means that the models can be well
constrained and tested by fitting to the majority of the event so that
they are fairly predictive for the unseen remainder of it.  Finally,
and very importantly, the low energy range has many problems
associated with it that are rather carefully treated in the dedicated
generators, such as exact kinematics, target mass effects, higher
order terms in the EPA and other polarization states of the photon.
This is in contrast with the QCD programs, which were traditionally
oriented towards the high energy limit and are only now starting to
`catch up' in their treatment of effects important at lower energy.

In conclusion, we would say that at pre-LEP energies existing programs
are perfectly adequate.  At LEP~1, they have proved sufficient for
most tasks, but problems are beginning to become apparent.  For
example, in DIS if one studies the energy flow in the detector as a
function of $x_{\mathrm{vis}},$ similar to what is shown in
Fig.~\ref{GGLLmckigney}, one obtains predictions that are reasonably
insensitive to the shape of the structure function and hence can make
a fairly stringent test of the model.  Preliminary results seem to
indicate that existing models already have
problems\cite{GGdelphiunfold}.  One can fairly confidently predict
that at LEP~2 these problems will be sufficiently severe that using
the QCD models will become essential, both because of the huge leap to
smaller $x$ and higher $Q^2$ and simply because of the higher
statistical precision that will require better control of systematic
errors.

Below we give a few basic facts about some of the programs used at
present.

\subsection{DIAG36}

\begin{tabbing}
{\bf Program location:} \= \kill
{\bf Author:}           \>  F.A.~Berends,~P.H. Daverveldt and R.~Kleiss \cite{GGSHDIAG36}\\
                        \> E-mail: t30@nikhef.nl         \\
{\bf Program size:}     \> 4335 lines                             \\
{\bf Program location:} \> CPC program library
\end{tabbing}
\GGdiag36\ is an event generator for the full set of QED diagrams for
$\mathrm{e^+e^- \to e^+e^-f\bar{f}},$ including all fermion masses.
Phase-space generation can cover the whole kinematically-allowed
phase-space, or cuts can be applied for typical single- or double-tagged
configurations.

\subsection{MINIJET}

\begin{tabbing}
{\bf Program location:} \= \kill
{\bf Author:}           \> A.~Miyamoto and H.~Hayashii \cite{GGAFtopazmc}               \\
                        \> E-mail:hayashii@naras1.kek.jp
\end{tabbing}
The \GGminijet\ program generates the direct and the resolved photon 
events in two-photon processes of $\mathrm{e^+e^-}$ collisions.
It calculates the cross-section of light- and heavy-quark production
according to the EPA approximation for the photon flux and the
leading-order matrix elements for the sub-process cross-sections.
It uses the BASES program for the cross-section calculation and
SPRING for the generation of four-momenta of final state partons.
The generated parton are hadronized using the \GGjetset\ program. 
It has been used in the analysis of two-photon data by the TOPAZ group
and in background studies for the JLC.
Multiple interactions are not included in the program.

\subsection{PC}

\begin{tabbing}
{\bf Program location:} \= \kill
{\bf Author:}           \>  F.L. Linde \cite{GGSHPC}\\
                        \>  E-mail: z66@nikhef.nl
\end{tabbing}
The \GGpc\ program generates two-photon events according to
Eq.~\ref{GGSHeq:lumfun}.  It provides event generation of fermion pairs
and resonance states in phase space of up to 4 final state particles
with many dominant resonant states.  The form factor is chosen between
$\rho, \pi,$ and $J/\psi$ poles.  It provides a cross-section
calculation of all terms in Eq.~\ref{GGSHeq:lumfun}.

\newpage
\subsection{RESPRO}

\begin{tabbing}
{\bf Program location:} \= \kill
{\bf Author:}           \> E.R.~Boudinov and M.V.~Shevlyagin \cite{GGSHRESPRO}\\
                        \>  E-mail: BOUDINOV@vxcern.cern.ch
\end{tabbing}
The \GGrespro\ program generates explicitly the two-photon resonance in
a very efficient way and calculates the cross-section according to
Eqs.~\ref{GGSHeq:lumfunTT} and~\ref{GGSHeq:BW}.

\subsection{TWOGAM}

\begin{tabbing}
{\bf Program location:} \= \kill
{\bf Authors:}          \> S.~Nova, A.~Olshevski, T.~Todorov \cite{GGLLtwogam}\\
                        \> E-mail: todorovt@vxcern.cern.ch
\end{tabbing}
\GGtwogam\ implements Eq.~\ref{GGSHeq:lumfun} exactly for leptonic final
states and for the direct, QPM, component of quark production, including
all helicity states and the exact $2\to4$ kinematics.  Single- and
double-resolved components are added to the transverse--transverse
scattering cross-section according to any parton distribution function
selected from \GGpdflib, and include a simple $P^2$-suppression model.
The resulting partonic states are hadronized by the Lund string model,
as implemented in \GGjetset.  The user can select from several different
soft VDM scattering models.

\subsection{TWOGEN}

\begin{tabbing}
{\bf Program location:} \= \kill
{\bf Author:}           \> A.~Buijs, W.G.J.~Langeveld, M.H.~Lehto,
                           D.J.~Miller \cite{GGLLtwogen}\\
                        \> E-mail: buijs@fys.ruu.nl          \\
{\bf Program size:}     \> 1800 lines                           \\
{\bf Program location:} \> http://www.fys.ruu.nl/~buijs/twogen/twogen.for
\end{tabbing}
\GGtwogen\ samples the transverse-transverse luminosity function for
real and virtual photons of the multiperipheral diagram.  The events
are then weighted with any user supplied cross-section
$\sigma_{TT}(\gamma\gamma\rightarrow X)$ and chosen using a simple
``hit or miss'' sampling. Thus, for example, the program can generate
according to a chosen $F_2$ or can produce resonances. Final state
partons are fragmented using the Lund string model in \GGjetset.

\newpage
\subsection{Vermaseren}

\begin{tabbing}
{\bf Program location:} \= \kill
{\bf Author:}           \> J.A.M.~Vermaseren  \cite{GGSHVermaseren}       \\
                        \> E-mail: t68@nikhef.nl
\end{tabbing}
The Vermaseren Monte Carlo has become the {\it de facto\/} standard
calculation of $\mathrm{e^+e^- \to e^+e^-f\bar{f}}$ via $\gamma\gamma$
collisions.  It uses the subset of the exact $2\to4$ matrix elements
in which the electron and positron lines do not annihilate (i.e.\
formally they apply to the process $\mathrm{e^+\mu^- \to
  e^+\mu^-f\bar{f}}$ with equal electron and ``muon'' masses).

\section{Conclusions}
\label{GGsummary}

$\gamma\gamma$ physics at LEP~2 is virtually impossible without event
generators. Not only are they needed for modelling detector effects, as
in almost all high energy physics analyses, but also because in
$\gamma\gamma$ collisions the energy of the incoming photons is
generally unknown.  This means that we need to reconstruct the basic
parameters of events solely from final-state properties.  Except for the
simplest final states consisting of only a few particles, detailed
understanding is presently only possible through models implemented in
event generators.
 
During the course of this workshop, several general purpose QCD
generators have been developed to better handle $\gamma\gamma$ and
e$\gamma$ interactions, enabling us to use experience gained from
experiments with $\mathrm{e^+e^-},$ ep and pp collisions, where these
programs have been used extensively. A lot more work needs to be done
however. In contrast to the special purpose generators used so far in
$\gamma\gamma$ physics, which in the case of high energy
multi-particle production are less theoretically advanced, the general
purpose ones have not yet been extensively confronted with available
data. Such comparisons have already been started for $\gamma$p data
from HERA with promising results, and it is important that this is
also done with LEP~1 $\gamma\gamma$ data in preparation for LEP~2.


\begin{thebibliography}{99}

\bibitem{GGLLggphys}
Report of the $\gamma\gamma$ physics working group, these proceedings.

\bibitem{GGLLqcdgen}
Report of the QCD event generator working group, these proceedings.

\bibitem{GGSHBudnev}
V.M.~Budnev, et al., \GGj{Phys.~Rep.} \GGv{15} (1975) 181.

\bibitem{GGSHBonneau}
G.~Bonneau, et al., \GGj{Nucl.~Phys.} \GGv{B54} (1973) 573.

\bibitem{GGSHField}
J.H.~Field, \GGj{Nucl.~Phys.} \GGv{B168} (1980) 477; \GGv{B176} (1980) 545;
\GGv{B187} (1981) 585.

\bibitem{GGLLtwogam}
S.~Nova, A.~Olshevski, T.~Todorov, DELPHI Note 90-35 (1990).

\bibitem{GGfrixione}
S.~Frixione, et al., \GGj{Phys.~Lett.} \GGv{B319} (1993) 339.


\bibitem{GGpdflib}
H.~Plothow-Besch, \GGj{Comput.~Phys.~Commun.} \GGv{75} (1993) 396.

\bibitem{GGRE-Schuler95a}
G.A.~Schuler and T.~Sj\"ostrand, \GGj{Z.~Phys.} \GGv{C68} (1995) 607.

\bibitem{GGRE-GRV92b}
M.~Gl\"uck, E.~Reya, A.~Vogt, \GGj{Phys.~Rev.} \GGv{D46} (1992) 1973.

\bibitem{GGLLstring}
B.~Andersson, et al., \GGj{Phys.~Rep.} \GGv{97} (1983) 31.

\bibitem{GGLLcluster}
B.R.~Webber, \GGj{Nucl.~Phys.} \GGv{B238} (1984) 492.

\bibitem{GGSHPC}
F.L.~Linde, \GGt{Charm production in two-photon collisions},
RX-1224 (LEIDEN), (1988) thesis.

\bibitem{GGSHDIAG36}
F.A.~Berends, P.H.~Daverveldt, R.~Kleiss, \GGj{Nucl.~Phys.} \GGv{B253} (1985) 421;
\GGj{Comput.~Phys.~Commun.} \GGv{40} (1986) 271, 285, and 309.

\bibitem{GGSHVermaseren}
J.A.M.~Vermaseren, \GGj{Nucl.~Phys.} \GGv{B229} (1983) 347.

\bibitem{GGSHPoppe}
M.~Poppe, \GGj{Int.~J.~of Mod.~Phys.} \GGv{1} (1986) 545.


\bibitem{GGSHLow} 
F.~Low, \GGj{Phys.~Rev.} \GGv{120} (1960) 582;
P.~Kessler, \GGj{Nuovo Cim.} \GGv{17} (1960) 809.

\bibitem{GGSHRESPRO}
E.R.~Boudinov and M.V.~Shevlyagin, DELPHI 95-49 PHYS 487.

\bibitem{GGLLtwogen}
A.~Buijs, et al., \GGj{Comput.~Phys.~Commun.} \GGv{79} (1994) 523.

\bibitem{GGRE-Thome77}
W.~Thom\'e, et al., \GGj{Nucl.~Phys.} \GGv{B129} (1977) 365.

\bibitem{GGLLblobel}
V.~Blobel, Proc.~CERN school of computing, Aiguablava, Spain, September 1984.

\bibitem{GGLLbayes}
G.~d'Agostini, \GGj{Nucl.~Instrum.~Meth.} \GGv{A362} (1995) 487. 

\bibitem{GGLLopal}
OPAL Collaboration, R.~Akers, et al., \GGj{Z.~Phys.} \GGv{C61} (1993) 199.

\bibitem{GGLLdelphi}
DELPHI Collaboration, P.~Abreu, et al., CERN-PPE/95-87, June 1995 (to be published in \GGj{Z.~Phys.~C}).

\bibitem{GGJacqBlon}
F.~Jacquet and A.~Blondel, \GGt{Proc.\ of the study for an ep facility
  for Europe}, DESY 79/48 (1979) 391.

\bibitem{GGzeusdijets}
ZEUS Collaboration, M.~Derrick, et al., \GGj{Phys.~Lett.} \GGv{B348} (1995) 665.

\bibitem{GGAFaleph}
ALEPH Collaboration, D.~Buskulic, et al., \GGj{Phys.~Lett.} \GGv{B355} (1995) 595.

\bibitem{GGAFdkzz}
M.~Drees, et al., \GGj{Phys.~Lett.} \GGv{B306} (1993) 371;\\
M.~Kr\"{a}mer, proceedings of the Photon 95 conference, Sheffield, UK.

\bibitem{GGAFtopazmc} A.~Miyamoto, H.~Hayashii,  KEK Preprint 94-204, to appear in \GGj{Comput.~Phys.~Commun.}

\bibitem{GGLLar}
L.~L{\"o}nnblad, \GGj{Comput.~Phys.~Commun.} \GGv{71} (1992) 15.

\bibitem{GGLLdcm1}
G.~Gustafson, \GGj{Phys.~Lett.} \GGv{B175} (1986) 453.

\bibitem{GGLLdcm2}
G.~Gustafson, U.~Pettersson, \GGj{Nucl.~Phys.} \GGv{B306} (1988) 746.

\bibitem{GGLLdcm3}
B.~Andersson, et al., \GGj{Z.~Phys.} \GGv{C43} (1989) 625.

\bibitem{GGAFBASES}
S.~Kawabata, \GGj{Comput.~Phys.~Commun.} \GGv{41} (1986) 127.

\bibitem{GGKEKjets}
T.~Munehisa, et al.,
KEK CP-034, KEK preprint 95-51, ENSLAPP-A-522/95, LPTHE-Orsay 95/37,
to appear in \GGj{Z.~Phys.~C}.

\bibitem{GGKEKll}
R.~Odorico, \GGj{Nucl.~Phys.} \GGv{B172} (1980) 157;\\
G.~Marchesini and B.R.~Webber, \GGj{Nucl.~Phys.} \GGv{B238} (1984) 1;\\
K.~Kato, T.~Munehisa, \GGj{Comput.~Phys.~Commun.} \GGv{64} (1991) 67.

\bibitem{GGKEKth}
E.~Witten, \GGj{Nucl.~Phys.} \GGv{B120}(1977)189;\\
Ch.~Berger and W.~Wagner, \GGj{Phys.~Rep.} \GGv{146} (1987) 1.

\bibitem{GGKEKksy}
K.~Kato, Y.~Shimizu, H.~Yamamoto, \GGj{Prog.~Theor.~Phys.} \GGv{63}(1980)1295;\\
K.~Kato, Y.~Shimizu, \GGj{Prog.~Theor.~Phys.} \GGv{64} (1980) 703;
\GGv{68} (1982) 862;\\
K.~Kato, Y.~Shimizu, H.~Yamamoto, preprint, UT-370 (1982), unpublished.

\bibitem{GGTSpyje}
T.~Sj\"ostrand, \GGj{Comput.~Phys.~Commun.} \GGv{82} (1994) 74; \\
T.~Sj\"ostrand, Lund University report LU TP 95--20 (1995).

\bibitem{GGMHSherwig}
G.~Marchesini, et al., \GGj{Comput.~Phys.~Commun.} \GGv{67} (1992) 465.

\bibitem{GGMHSsmallx}
G.~Marchesini and B.R.~Webber, \GGj{Nucl.~Phys.} \GGv{B386} (1992) 215.

\bibitem{GGMHShard}
M.H.~Seymour, \GGj{Comput.~Phys.~Commun.} \GGv{90} (1995) 95. 

\bibitem{GGMHSglasgow}
M.H.~Seymour, contribution gls0258 to the
\GGt{27th International Conference on High Energy Physics},
Glasgow, U.K., 20--27 July 1994.

\bibitem{GGMHSjim}
J.M.~Butterworth, J.R.~Forshaw, \GGj{J.~Phys.} \GGv{G19} (1993) 1657;\\
J.M.~Butterworth, J.R.~Forshaw, M.H.~Seymour, CERN--TH/95--82, in preparation.

\bibitem{GGRE-Engel95a}
R.~Engel, \GGj{Z.~Phys.} \GGv{C66} (1995) 203;\\
R.~Engel.
\GGt{Multiparticle Photoproduction within the two-component Dual Parton Model},
in preparation,  1995;\\
R.~Engel, J.~Ranft, \GGt{Hadronic photon-photon collisions at high energies},
ENSLAPP-A-540/95 (hep-ph/9509373), 1995.

\bibitem{GGRE-Capella94a}
A.~Capella, et al., \GGj{Phys.~Rep.} \GGv{236} (1994) 227.

\bibitem{GGRE-Aurenche92a}
P.~Aurenche, et al., \GGj{Phys.~Rev.} \GGv{D45} (1992) 92.

\bibitem{GGRE-Aurenche94a}
P.~Aurenche, et al., \GGj{Comput.~Phys.~Commun.} \GGv{83} (1994) 107.

\bibitem{GGRE-Abramovski73}
V.A.~Abramovski, V.N.~Gribov, O.V.~Kancheli, \GGj{Yad.~Fis.} \GGv{18} (1973) 595.

\bibitem{GGRE-Hahn90}
K.~Hahn and J.~Ranft, \GGj{Phys.~Rev.} \GGv{D41} (1990) 1463.

\bibitem{GGRE-Capella80b}
A.~Capella, et al., \GGj{Z.~Phys.} \GGv{C10} (1980) 249;\\
A.B.~Kaidalov, \GGj{Phys.~Lett.} \GGv{B116} (1982) 459.

\bibitem{GGRE-Capella95a}
A.~Capella, et al., \GGj{Phys.~Lett.} \GGv{B343} (1995) 403;\\
R.~Engel, J.~Ranft, S.~Roesler, \GGv{Phys.~Rev.} \GGv{D52} (1995) 1459.

\bibitem{GGRE-Bengtsson84}
H.U.~Bengtsson, \GGj{Comput.~Phys.~Commun.} \GGv{31} (1984) 323.

\bibitem{GGRE-Engel95e}
R.~Engel, \GGphojet\ manual, 1995.

\bibitem{GGTSgamp}
G.A.~Schuler, T.~Sj\"ostrand, \GGj{Phys.~Lett.} \GGv{B300} (1993) 169;
\GGj{Nucl.~Phys.} \GGv{B407} (1993) 539.

\bibitem{GGTSgamgam}
G.A.~Schuler and T.~Sj\"ostrand, in
\GGt{Two-Photon Physics from DA$\Phi$NE to LEP200 and Beyond},
World Scientific, Singapore, 1994, eds.~F.~Kapusta and J.~Parisi, p163; \\
T.~Sj\"ostrand, in
\GGt{XXIV International Symposium on Multiparticle Dynamics 1994},
World Scientific, Singapore, 1995, eds.~A.~Giovannini, S.~Lupia and
R.~Ugoccioni, p221.

\bibitem{GGTScross}
G.A.~Schuler and T.~Sj\"ostrand, \GGj{Phys.~Rev.} \GGv{D49} (1994) 2257.

\bibitem{GGMHSh1}
H1 Collaboration, I.~Abt, et al., \GGj{Z.~Phys.} \GGv{63} (1994) 377.

\bibitem{GGLLFKP}
J.H.~Field, F.~Kapusta, L~Poggioli, \GGj{Phys.~Lett.} \GGv{B181} (1986) 362;
J.H.~Field, F.~Kapusta, L~Poggioli, \GGj{Z.~Phys.} \GGv{C36} (1987) 121;
F.~Kapusta, \GGj{Z.~Phys.} \GGv{C42} (1989) 225.

\bibitem{GGdelphiunfold}
F.~Kapusta, I.~Tyapkin, N.~Zimin and A.~Zinchenko, DELPHI note in preparation.

\end{thebibliography}
\end{document}